\documentclass[journal]{IEEEtran}

\ifCLASSINFOpdf

\else

\fi
\usepackage{comment}
\usepackage[utf8]{inputenc}
\usepackage{amsmath}
\usepackage{float}
\usepackage{mathtools}
\usepackage{braket}
\usepackage{graphicx}
\graphicspath{{./figures/}}
\usepackage{amsfonts}
\usepackage[ruled,vlined,linesnumbered]{algorithm2e}
\let\oldnl\nl% Store \nl in \oldnl
\newcommand{\nonl}{\renewcommand{\nl}{\let\nl\oldnl}}% Remove line number for one line

\usepackage{soul}
\usepackage{csquotes}
\usepackage{hhline}
\usepackage{amssymb}
\usepackage{listings}
\usepackage{color}
\definecolor{codegreen}{rgb}{0,0.6,0}
\definecolor{codegray}{rgb}{0.5,0.5,0.5}
\definecolor{codepurple}{rgb}{0.58,0,0.82}
\definecolor{backcolour}{rgb}{0.95,0.95,0.92}

\usepackage{subcaption}
\usepackage{titlesec}

\usepackage{dcolumn}
\usepackage{tabularx}
\usepackage{hyperref}
\usepackage{amsmath}

\usepackage{algorithmic}
\usepackage{graphicx}

\hypersetup{
    colorlinks=true,
    linktoc=all,
    linkcolor=black,
    citecolor=magenta,
}
\usepackage{longtable}
\usepackage{braket}
\newcolumntype{C}{>{\centering\arraybackslash}X}
\SetAlFnt{\small}
\usepackage[font=small]{caption}
\begin{document}
% correct bad hyphenation here
\hyphenation{op-tical net-works semi-conduc-tor}

\title{QFDNN: A Resource-Efficient Variational Quantum Feature Deep Neural Networks for Fraud Detection and Loan Prediction}

%\title{Analysis of $UU^{\dagger}$, Variational $UU^{\dagger}$ and Quantum Neural Network Methods in the Presence of Noise for Classification Tasks in the IoT Extreme Environment}

%\author{Sritam Kumar Satpathy\thanks{S.~K.~Satpathy is with the Department of Physical Sciences, Indian Institute of Science Education and Research, Berhampur, India e-mail: (sritamkumar04@gmail.com).}, Vallabh Vibhu\thanks{V.~Vibhu is with the Department of Physical Sciences, Indian Institute of Science Education and Research, Berhampur, India e-mail: (vallabhvibhu07@gmail.com).}, Bikash~K.~Behera\thanks{B.~K. Behera is with the Bikash's Quantum (OPC) Pvt. Ltd., Mohanpur, WB, 741246 India, e-mail: (bikas.riki@gmail.com).}, Saif Al-Kuwari\thanks{Saif~Al-Kuwari is with the College of Science and Engineering, Hamad Bin Khalifa University, Qatar Foundation, Doha, Qatar. e-mail: (smalkuwari@hbku.edu.qa).}, Shahid Mumtaz\thanks{Shahid~Mumtaz is with the Department of Applied Informatics Silesian University of Technology Akademicka 16 44-100 Gliwice, Poland, and Nottingham Trent University, Engineering Department, United Kingdom. e-mail: (dr.shahid.mumtaz@ieee.org).} and Ahmed~Farouk\thanks{A.~Farouk is with the Department of Computer Science, Faculty of Computers and Artificial Intelligence, South Valley University, Hurghada, Egypt. e-mail: (ahmed.farouk@sci.svu.edu.eg).}}%
%\markboth{\today}%
\author{Subham~Das, Ashtakala Meghanath, Bikash~K.~Behera,  Shahid~Mumtaz, Saif~Al-Kuwari, Ahmed~Farouk
\thanks{Subham Das is with the Department of Physics, Indian Institute of Science Education and Research, Thiruvananthapuram, India; Email: Subhamdas19@alumni.iisertvm.ac.in}
\thanks{Ashtakala Meghanath is with the Department of Physics, Indian Institute of Science Education and Research, Thiruvananthapuram, India;  Email: meghanath19@alumni.iisertvm.ac.in}
\thanks{B.~K. Behera is with Bikash's Quantum (OPC) Pvt. Ltd., Mohanpur, WB, 741246 India; Email: bikas.riki@gmail.com}
\thanks{Shahid~Mumtaz is with Nottingham Trent University, Engineering Department, United Kingdom. Email: (dr.shahid.mumtaz@ieee.org).}
\thanks{S.~Al-Kuwari is with the Qatar Center for Quantum Computing, College of Science and Engineering, Hamad Bin Khalifa University, Doha, Qatar. e-mail: (smalkuwari@hbku.edu.qa).}
\thanks{A.~Farouk is with the Qatar Center for Quantum Computing, College of Science and Engineering, Hamad Bin Khalifa University, Doha, Qatar and with the Department of Computer Science, Faculty of Computers and Artificial Intelligence, Hurghada University, Hurghada, Egypt; Email: ahmedfarouk@ieee.org}

\thanks{\textit{Corresponding Authors: Ahmed~Farouk}}
}

\maketitle
\begin{abstract}
Social financial technology focuses on trust, sustainability, and social responsibility, which require advanced technologies to address complex financial tasks in the digital era. With the rapid growth in online transactions, automating credit card fraud detection and loan eligibility prediction has become increasingly challenging. Classical machine learning (ML) models have been used to solve these challenges; however, these approaches often encounter scalability, overfitting, and high computational costs due to complexity and high-dimensional financial data.  Quantum computing (QC) and quantum machine learning (QML) provide a promising solution to efficiently processing high-dimensional datasets and enabling real-time identification of subtle fraud patterns. However, existing quantum algorithms lack robustness in noisy environments and fail to optimize performance with reduced feature sets. To address these limitations, we propose a quantum feature deep neural network (QFDNN), a novel, resource efficient, and noise-resilient quantum model that optimizes feature representation while requiring fewer qubits and simpler variational circuits. The model is evaluated using credit card fraud detection and loan eligibility prediction datasets, achieving competitive accuracies of 82.2\% and 74.4\%, respectively, with reduced computational overhead. Furthermore, we test QFDNN against six noise models, demonstrating its robustness across various error conditions. Our findings highlight QFDNN’s potential to enhance trust and security in social financial technology by accurately detecting fraudulent transactions while supporting sustainability through its resource-efficient design and minimal computational overhead.
\end{abstract}

\IEEEpeerreviewmaketitle

\section{Introduction}\label{QVP:Sec1}

Financial decision-making is critical in today's rapid social digital landscape, particularly within the banking and financial services sector. The increasing volume of loan applications, online transactions, and digital financial activities led to significant challenges, such as transaction fraud \cite{8428484, ebiaredoh2021artificial}. Furthermore, the complexity and time constraints of processing financial data and learning from that data are difficult and exceed the capabilities of traditional linear and heuristic models. Therefore, it is essential to develop advanced models capable of automating complex financial tasks and making more accurate and reliable decisions \cite{Hawley1990}.

Deep Neural Networks (DNNs) have been identified as a powerful tool that can be used to address complex financial analysis problems (FAPs) in different domains such as risk management \cite{10491318}, fraud detection, portfolio analysis \cite{10242155}, and market prediction \cite{10613029}. DNNs can analyze historical data and discover trends to achieve accurate predictions using advanced machine learning (ML) techniques such as reinforcement learning (RL) \cite{TAN2024102049}. Various applications have been demonstrated in tasks such as credit scoring \cite{KWON2025125327}, algorithmic trading \cite{GRUDNIEWICZ2023102052}, and customer segmentation, which can be beneficial to better decision-making and operational efficiency of financial institutions. DNNs have also shown significant potential in quantitative finance by modeling complex financial instruments such as option pricing \cite{culkin2017machine} and solving stock market prediction problems \cite{SNASEL2024102018} and investment strategy \cite{sen2022introductory}. In addition, DNNs have been implemented to improve the Brazil inflation forecast \cite{ARAUJO2023100087}, efficient average pricing options based on \cite{GAN2020119928}, and implied volatility surface (IVS) modeling with adaptive support vector regression (SVR) \cite{ZENG2019376}. Models have been proposed that can enhance credit card fraud detection by learning behavioural patterns from users' historical transactions from long and short-term habits \cite{9912385,9744717} and remove noisy data from imbalanced datasets \cite{10063201}.

Classical ML and DNNs suffer from multiple shortcomings that limit their predictive and decision-making capabilities in financial problems \cite{10319418}. The financial data often has inconsistent quality and quantity that can create problems with noise, incompleteness, and bias that affect model performance. If the data is insufficient, then the model can be overfitted, reducing its ability to generalize to new scenarios.
Despite these advancements, feature selection remains a critical challenge in DNN-based financial models, as poorly selected features can lead to biased or inaccurate predictions \cite{9738619}. Moreover, DNNs often struggle to fully understand the inherent risks in financial decisions, resulting in algorithmic bias and challenges in regulatory compliance \cite{10433778}. Furthermore, integrating DNNs into existing financial infrastructures requires extensive modifications, adding complexity to the deployment \cite{THAKKAR2021114800}. Therefore, addressing these challenges is essential to enhance the robustness, accuracy, and scalability of financial decision-making models.

Quantum computing (QC) combined with FAPs introduces a novel category called quantum financial analysis problems (QFAPs) to enhance decision-making, prediction, and anomaly detection. Traditional DNN models have significantly advanced financial analysis by automating stock market prediction, risk assessment, and fraud detection processes. However, the complexity and scale of modern financial data exceed the capabilities of classical computational resources. Thus, with its unique properties, such as entanglement and superposition, QC presents encouraging advantages over classical computing in processing and analyzing large datasets and solving complex problems efficiently. Quantum algorithms such as Shor's algorithm \cite{ugwuishiwu2020overview} and Deutsch-Jozsa algorithm have shown outperforming performance compared to classical in various impactful applications \cite{Nagata2017}. Furthermore, quantum machine learning (QML) has been shown to surpass the best classical ML approaches \cite{Du2022}. Quantum-enhanced models, such as quantum deep neural networks (QDNNs), can deliver faster optimization \cite{Beer2020} and more accurate predictions, especially when dealing with multi-feature financial datasets. 

QDNNs represent an innovative fusion of QC with classical DNNs, allowing the analysis and prediction of complex financial data with increased efficiency and accuracy \cite{Orus2019}. This approach utilizes the quantum mechanical framework, which can be derived naturally from Fisher information \cite{NASTASIUK20151998}, offering a resource-efficient and time-saving method for QDNNs to learn complex patterns in the data without the risk of overfitting. QDNNs are well-suited for dynamic financial environments where market conditions and trends change rapidly. Several quantum algorithms have been developed to address specific financial decision-making challenges. For example, Hybrid Quantum Neural Networks (HQNNs) integrate the inherent parallelism of QC with classical neural networks to improve feature representation and learning efficiency \cite{PAQUET2022116583}. Similarly, Quantum Graph Neural Networks (QGNNs) utilize graph-based structures to analyze relationships within financial data, capturing complex temporal patterns and enabling more accurate predictions \cite{10545170}. In addition, the matrix-based approach has become a new method for encoding financial time series into quantum states. Using quantum encoders coupled with classical measurement networks \cite{PAQUET2022116583}, these methods can predict future states with high fidelity. Furthermore, false-positive and false-negative optimization is included in the ensemble of quantum methods used to assess credit risk through classification and regression algorithms to increase precision in predicting default probabilities and exposure levels \cite{PAPOUSKOVA201933}. These advances highlight the potential impact of quantum algorithms in automating and improving financial decision making.

However, existing classical ML has several limitations, including overfitting, inconsistent data quality, noise, incompleteness, and bias, which affect performance. Overfitting is a common challenge, particularly when data are limited, which requires frequent retraining to adapt to the dynamic nature of financial markets \cite{ai5040101}. Furthermore, current quantum models are vulnerable to noise, which compromises system stability and reduces prediction reliability. These models also require significant quantum resources, particularly qubits, making them challenging to implement on current quantum hardware. 
In addition, the use of large variational layers \cite{McClean2018}, time-consuming \cite{PhysRevLett.128.180505}, and resource-inefficient optimization techniques results in significant training costs and time. In addition, inefficient feature reduction and optimization processes cause resource constraints, limiting the scalability and practicality of quantum algorithms for high-dimensional financial datasets. Overcoming these challenges is essential for developing robust and efficient financial decision-making systems.

Therefore, we introduce the Quantum Feature Deep Neural Network (QFDNN) as a resource-efficient solution for financial decision-making, specifically credit card fraud detection and loan eligibility prediction. QFDNN leverages the UU† method and variational quantum circuits (VQC) to enhance feature learning, reducing overfitting and computational overhead compared to classical methods. It achieves competitive accuracy of 82.2\% and 74.4\% in fraud detection and loan prediction datasets while demonstrating resilience to noise models such as bit flip and phase flip, ensuring reliability in noisy quantum environments. Existing quantum models, such as LEP-QNN and FHQNN, require a large number of qubits, limiting their practicality for near-term quantum hardware. In contrast, QFDNN introduces an optimized multiple-qubit $UU^{\dagger}$ encoding scheme, allowing exponential compression of features, encoding $2^N-1$ features using only $N$ qubits, making it highly scalable and feasible for NISQ-era devices. Additionally, QFDNN employs COBYLA optimization within VQCs, ensuring stable training while minimizing computational overhead, unlike deeper quantum models requiring extensive qubit connectivity. By addressing the scalability, resource constraints, and noise challenges of quantum computing, QFDNN is a practical and efficient alternative to classical and existing quantum models, making it a promising candidate for real-world financial applications.

%\subsection{Novelty and contribution}
The contributions of this paper can be summarized as follows:
\begin{itemize}
\item[1)] Propose a novel resource-efficient, multi-feature and noise-resilient quantum feature deep neural network (QFDNN) model that leverages the $UU^\dagger$ method and the variational quantum neural network (VQNN) for financial decision-making tasks.
\item[2)] Demonstrate the efficiency and performance of QFDNN through simulation experiments on two financial datasets, achieving competitive accuracy and resource efficiency compared to classical and quantum machine learning algorithms.
\item[3)] Verify the robustness of QFDNN against six different noise models, showcasing its resilience in noisy quantum environments, which is essential for real-world applications of QC in finance.
\end{itemize}

The remainder of the paper is organized as follows. Section \ref{SecII} reviews the related work, followed by Section \ref{SecIII}, where the methodology of the proposed QFDNN model is described. Section \ref{SecIV} presents the experimental results, comparing QFDNN's performance with classical and other quantum models across two financial datasets. Finally, Section \ref{SecV} concludes the paper with a discussion of the findings.

\begin{figure*}[]
\centering
\includegraphics[width=\linewidth]{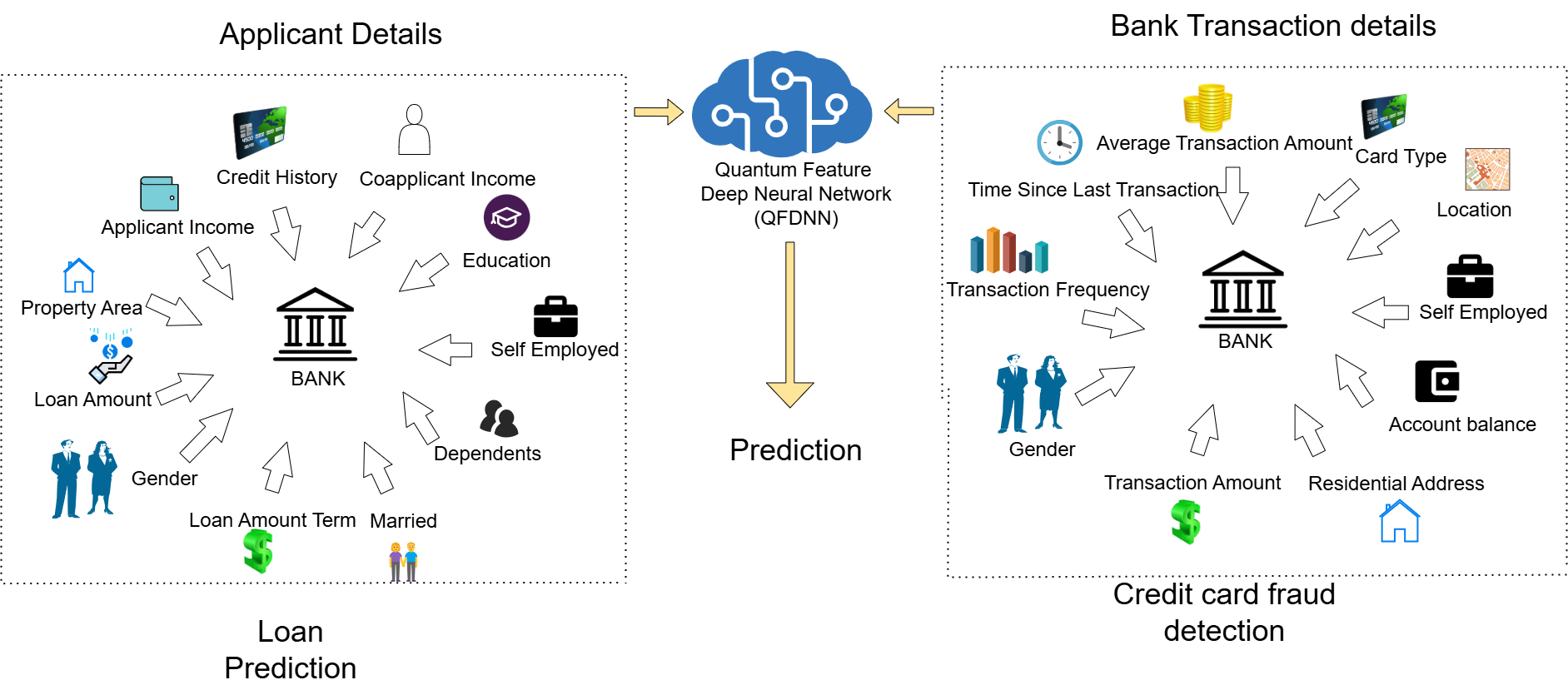}
\caption{Schematic Representation of the QFDNN's Application}
\label{Schematic fig}
\end{figure*}
%%%%%%%%%%%%%%%%%%%%%%%%%%%%%%%%%%%%%%%%%%%%%%%%%%%%%%
\section{Related Work \label{SecII}}
Classical ML and DL approaches are applied to solve complex financial decision-making problems.
For example, ensemble learning techniques combined with feature engineering have demonstrated improved accuracy in credit card fraud detection by enhancing feature representation and reducing redundancy\cite{FraudDetection2022}. Similarly, stacked sparse autoencoder-based artificial neural networks have been employed to improve credit scoring predictions, addressing issues of feature complexity and scalability \cite{CreditScoring2021}.
A fully connected feed-forward neural network to replicate the Black-Scholes option pricing formula with high accuracy, highlighting the promise of DL in financial modelling and the importance of hyperparameter tuning, is presented in \cite{culkin2017machine}. The roles of AI and ML applications in banking, such as credit scoring and fraud detection, while addressing challenges related to data interpretability and system integration, are examined in \cite{sen2022introductory}. 
Meanwhile, \cite{GAN2020119928} utilized ML to price average options, reducing repetitive computations efficiently; however, additional validation is required for its application to more complex financial products. An adaptive SVR model for implied volatility surfaces that shows impressive computational speedup but is restricted by the need for advanced hardware, which limits broader accessibility, was proposed in \cite{ZENG2019376}. Furthermore, in \cite{PAPOUSKOVA201933}, a two-stage ensemble model was proposed for credit risk prediction, achieving improved accuracy but at the cost of increased computational complexity. Integrating dynamic transaction behaviors into a Radial Basis Function framework improved credit risk assessment and resulted in better predictions; however, it required extensive feature engineering \cite{ZHANG201865}. Filter-based feature selection techniques have been shown to improve model accuracy for chronic disease prediction, demonstrating how feature selection can significantly impact model performance in complex datasets \cite {FeatureSelection2022}, and autoencoder-based feature extraction has proven effective in improving classification tasks in healthcare \cite {DiseaseDetection2020}. The expanding objectives of ML and DL in financial decision-making continue to focus on addressing persistent challenges such as scalability, computational resource demands, overfitting, and efficient handling of complex high-dimensional data, as discussed in \cite{ai5040101}.
 
Several quantum approaches, such as the Loan Eligibility Prediction Quantum Neural Network (LEP-QNN), achieved impressive accuracy by incorporating a dropout mechanism to reduce overfitting and improve predictive reliability \cite{innan2024lepqnnloaneligibilityprediction}. However, scalability and robustness to quantum noise remain challenges for practical deployment. An HQNN utilizing density matrices to model financial time series and predict maximum prices is proposed in \cite{PAQUET2022116583}; however, the system proved efficient in parallelized learning, and its applicability to large-scale datasets requires further validation. Similarly, QGNN for fraud detection surpassed traditional methods despite challenges in scalability and noise resistance \cite{Innan_2024}. A deep quantum neural network applied to option pricing and implied volatility shows promise for numerical financial problems, but needs broader validation using larger datasets \cite{sakuma2022applicationdeepquantumneural}. A hybrid quantum-classical credit scoring model was introduced, demonstrating faster training times than classical methods, although it encountered scalability challenges \cite{math12091391}. 
To overcome these limitations, we propose a QFDNN as a resource-efficient and noise-resilient model for social fintech applications, including tasks such as predicting loan eligibility and detecting credit card fraud (see Fig. \ref{Schematic fig}). Circuit optimization is achieved using a variational method alongside the COBYLA (Constrained Optimization BY Linear Approximations) optimizer, ensuring effective parameter tuning. 

\begin{figure*}[]
\centering
\includegraphics[width=\linewidth]{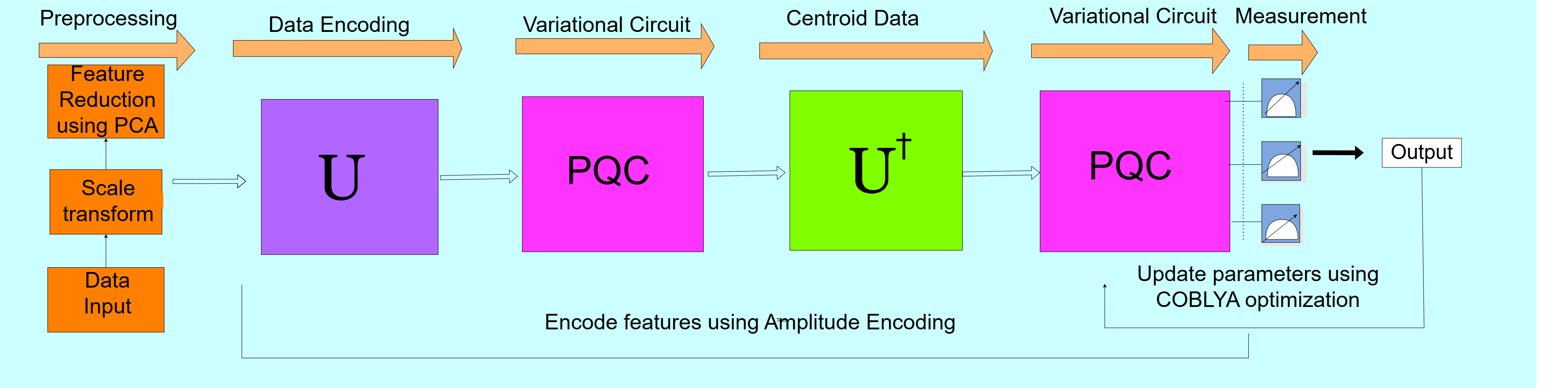}
\caption{QFDNN System Model.}
\label{fig_1}
\end{figure*}
%\newpage
%\newpage
\begin{figure*}[]
\centering
\begin{subfigure}{.5\textwidth}
    \includegraphics[width=\linewidth]{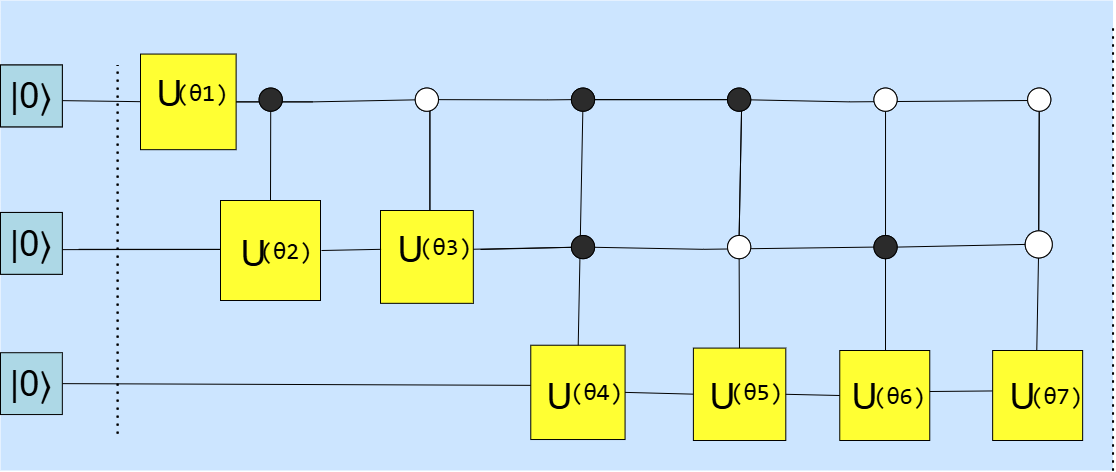}
    %{Circuitdiagram/}
    \caption{}
    \label{Encode}
\end{subfigure}\hfill
\begin{subfigure}{.5\textwidth}
    \includegraphics[width=\linewidth]{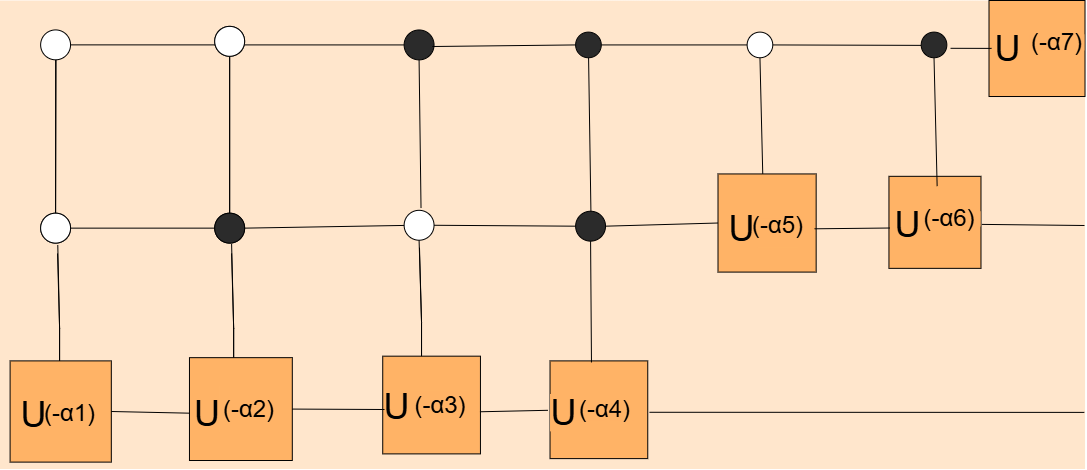}

    %%{Circuitdiagram/Copy of Untitled Diagram.drawio (2).png}
    \caption{}
    \label{centroid}
\end{subfigure}\hfill
\begin{subfigure}{.5\textwidth}
    \includegraphics[width=\linewidth]{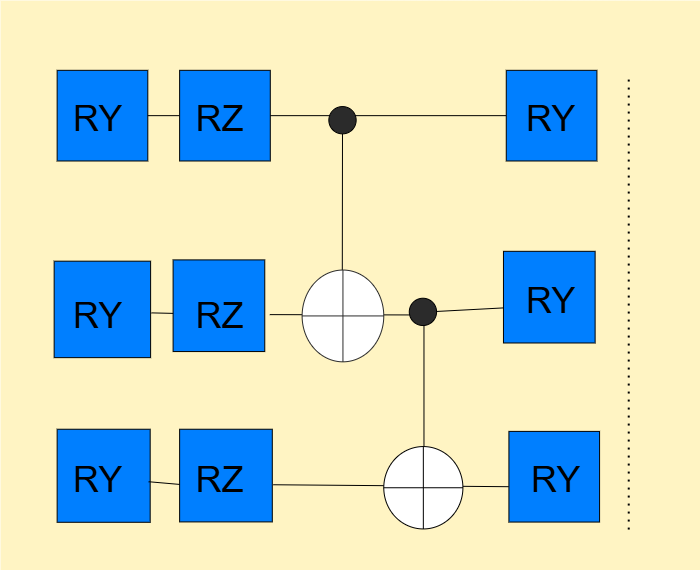}
    \caption{}
    \label{variational}
\end{subfigure}\hfill
\caption{Data Encoding for (a) QFDNN; (b) Centroid Data; (c) Variational Part.}
\label{fig_21}
\end{figure*}

%%%%%%%%%%%%%%%%%%%%%%%%%%%%%%%%%%%%%%%%%%%%%%%%%%%%%%
\section{Methodology \label{SecIII}}
The QFDNN development process follows a structured sequence of stages as shown in Fig. \ref{fig_1}. The first stage is data preprocessing, which involves applying Principal Component Analysis (PCA) to reduce the dimensionality and scaling the feature values for consistency. This is followed by a multiple-qubit amplitude encoding strategy to efficiently represent the data within a quantum circuit. $UU^\dagger$ is the main method that quantifies how closely the dataset approximates the ideal case and aids effective feature representation. Parameterized quantum circuits (PQCs) are integrated at two points after the $U$ and $U^\dagger$ circuit. These PQCs are optimized using a classical optimizer and applied to test data points for classification and evaluation. Finally, the QFDNN's robustness is validated against various noise models to ensure its resilience in real-world noisy environments.

\subsection{$UU^{\dagger}$ Method}
The $UU^\dagger$ method is a fundamental component in the QFDNN architecture, as it computes the inner product between the quantum states that represent the centroid points \(|C\rangle\) and the test data \(|T\rangle\). It enables QFDNN to assess how closely the data aligns with an ideal representation to achieve accurate classification and decision-making in financial tasks such as fraud detection and loan prediction. The $UU^\dagger$ process is as follows:

%As mentioned earlier, the $UU^\dagger$ method is a a fundamental component in the architecture of QFDNN, computing the inner product between quantum states representing the centroid \(|C\rangle\) and test data \(|T\rangle\) points. This method enables the QFDNN to assess how closely the data aligns with an ideal representation to achieve accurate classification and decision-making in financial tasks such as fraud detection and loan prediction. By utilizing the $UU^\dagger$ method, QFDNN effectively captures complex relationships within the data, forming the foundation for its robust performance in noiseless and noisy quantum environments.  The process involves:

\begin{itemize} 

    \item \textbf{State Preparation}: Encode centroid data \(|C\rangle\) using unitary \(U_1\) and test data \(|T\rangle\) using unitary \(U_2\):
   \begin{equation}
    \ket{C} = U_1 \ket{0}^{\otimes n}, \quad \ket{T} = U_2 \ket{0}^{\otimes n}.
\end{equation}

    \item \textbf{Operator Definition}: Define the operator \(A\) as: \begin{equation}
         A = U_2^\dagger U_1
    \end{equation}

    \item \textbf{Inner Product}: Calculate the inner product \(\langle C|T \rangle\) by applying \(A\) to \(|0\rangle^{\otimes n}\) and measuring the amplitude \(a_0\) of the state \(|0\rangle^{\otimes n}\):
   \begin{equation}
    \braket{C|T} = \sqrt{P_{\ket{0}^{\otimes n}}}, \quad P_{\ket{0}^{\otimes n}} = |a_0|^2.
\end{equation}

\end{itemize}

This requires implementing quantum circuits for \(U_1\), \(U_2\), and measurements in a quantum computation framework (see Algorithm \ref{UUalgo}).

\iffalse
\begin{algorithm}
\DontPrintSemicolon
\caption{$UU^\dagger$ Method}
\label{alg:UU_method}

\SetKwFunction{CircuitFunction}{Circuit\_function}
\SetKwProg{Fn}{def}{:}{end}
\Fn{\CircuitFunction{$params$}}{
    \textbf{Apply unitary gates:}\;
    Initialize quantum state to $\ket{0_{1}, 0_{2}, \ldots, 0_{n}}$\;
    
    Apply unitary gate $U(\theta_{1}, \theta_{2}, \ldots, \theta_{n})$ to the quantum state:\;
    $\ket{\psi} \gets U(\theta_{1}, \theta_{2}, \ldots, \theta_{n}) \ket{0_{1}, 0_{2}, \ldots, 0_{n}}$\;
    
    Apply unitary gate $U(-\theta'_{1}, -\theta'_{2}, \ldots, -\theta'_{n})$ to the quantum state:\;
    $\ket{\psi'} \gets U(-\theta'_{1}, -\theta'_{2}, \ldots, -\theta'_{n}) \ket{\psi}$\;
    
    \textbf{Perform measurement:}\;
    Measure the circuit in the computational basis.\;

    \textbf{Calculate inner product:}\;
    Compute the inner product between the two measurement outcomes by taking the square of the probability of $\ket{0_{1}, 0_{2}, \ldots, 0_{n}}$\;

    \textbf{Classification:}\;
    Apply the classification condition to the inner product.\;

    \textbf{Calculate accuracy:}\;
    Compute the accuracy of the classification.\;
}
\end{algorithm}
\fi

\begin{algorithm}
\DontPrintSemicolon
\caption{$UU^\dagger$ Method}
\label{UUalgo}

\nonl \textbf{Input:} Normalized cleaned and scaled dataset\;
\nonl \textbf{Output:} accuracy\;

\SetKwFunction{CircuitFunction}{Circuit\_function}
\SetKwProg{Fn}{def}{:}{end}
\Fn{\CircuitFunction{$params$}}{
    \nonl \textbf{Apply unitary gates}\;
    \nonl Initialize quantum state into $\ket{0_{1},0_{2},...0_{n}}$\;
    
    \nonl Apply unitary gate $U(\theta_{1},\theta_{2},..\theta_{n})$ to the quantum state: $\ket{\psi} \gets U(\theta_{1},\theta_{2},..\theta_{n}) \ket{0_{1},0_{2},...0_{n}}$\;
    
    \nonl Apply unitary gate $U(-\theta'_{1},-\theta_{2},..-\theta_{n})$ to the quantum state: $\ket{\psi'} \gets U(-\theta'_{1},-\theta_{2},..-\theta_{n}) \ket{\psi}$\;
    
    \textbf{Perform Measurement}\;
    \nonl Measure the circuit on the computational basis.\;
    
    \textbf{Calculate inner product}\;
    \nonl Compute the inner product between the two measurement outcomes by taking the square of the probability of $\ket{0_{1},0_{2},...0_{n}}$\;

    \textbf{Classification}\;
    \nonl The classification condition on the inner product is applied\;
    
    \textbf{Accuracy}\;
    \nonl The accuracy is calculated\;
}

\end{algorithm}

%%% changed this 

% \begin{algorithm}
% \DontPrintSemicolon
% \caption{$UU^\dagger$ method}
% \label{UU algo}

% \nonl\textbf{Input:}  Normalized cleaned and scaled dataset\;
% \newline
% \nonl\textbf{Output:} accuracy \;
    
% \SetKwFunction{CircuitFunction}{Circuit\_function}
% \SetKwProg{Fn}{def}{:}{end}
% \Fn{\CircuitFunction{$params$}}{
%     \nonl\textbf{Apply unitary gates}\;
%     \nonl    Initialize quantum state into $\ket{0_{1},0_{2},...0_{n}}$\;
%     \newline
% \nonl    Apply unitary gate $U(\theta_{1},\theta_{2},..\theta_{n})$ to the quantum state: $\ket{\psi} \gets U(\theta_{1},\theta_{2},..\theta_{n}) \ket{0_{1},0_{2},...0_{n}}$\;
% \newline
% \nonl    Apply unitary gate $U(-\theta'_{1},-\theta_{2},..-\theta_{n})$ to the quantum state: $\ket{\psi'} \gets U(-\theta'_{1},-\theta_{2},..-\theta_{n}) \ket{\psi}$\;
% \newline
%     \textbf{Perform Measurement}\;
% \nonl    Measure the circuit in the computational basis.\;

%     \textbf{Calculate inner product}\;
% \nonl    Compute the inner product between the two measurement outcomes by taking the square of the probability of $\ket{0_{1},0_{2},...0_{n}}$\;

%  \textbf{Classification}\;
    
% \nonl  The classification condition on the inner product is applied\;
%     \textbf{Accuracy}\;
% \nonl The accuracy is calculated 
    
% }

% \end{algorithm}

\subsection{Multiple Qubit Amplitude Encoding}
Here, the multiple-qubit amplitude encoding technique is designed to effectively encode a set of financial features into a quantum circuit as part of the QFDNN model (Fig. \ref{fig_21}). The encoding strategy is as follows:
\begin{enumerate}
\item First Qubit: A single $U$ gate operation encodes the first feature.
\item Second Qubit: A controlled $U$ gate and an anti-controlled $U$ gate are applied, with the first qubit acting as the controller to represent interdependencies between the first and second features.
\item For the Third Qubit: A $U$ gate is applied with the first and second qubits as controllers. Another $U$ gate is also implemented with the first qubit as a controller and the second as the anti-controller. A third $U$ gate involves the first qubit as the anti-controller and the second as the controller. A fourth $U$ gate is applied with both the first and second qubits as an anti-controller. 
\item Fourth Qubit: Eight $U$ gates are applied, covering all possible controller and anti-controller combinations among the first three qubits to capture higher-order interactions between features.
\item Higher Qubits: Following similar logic, the process is extended iteratively for the fifth qubit and beyond to encode additional features, ensuring scalability for high-dimensional financial datasets.
   % \item On the first qubit, a single $U$ gate operation is performed.
    %\item On the second qubit, a controlled $U$ gate and an anti-control $U$ gate are performed, where the first qubit serves as the controller.
    %\item For the third qubit, a $U$ gate is applied with both the first and second qubits as controllers. Additionally, another $U$ gate is implemented with the first qubit as the controller and the second qubit as the anti-controller. A third $U$ gate involves the first qubit as the anti-controller and the second qubit as the controller. A fourth $U$ gate is applied with both the first and second qubit as an anti-controller. 
    %\item The fourth qubit undergoes eight $U$ gates, covering all possible combinations of controllers and anti-controllers among the first three qubits.
   % \item The process is iterated for the fifth qubit and any other higher number of qubits.
\end{enumerate}
This encoding strategy combines single-qubit gates $U$ and controlled operations to transform classical financial data into quantum states in a multiqubit system within the QFDNN architecture (see Algorithm \ref{alg:amplitude_encoding} and Figures \ref{Encode} and \ref{centroid}). In the $U$ circuit, the feature values are encoded sequentially, while in the $U^{\dagger}$ circuit, they are encoded in reverse order. It uses a logarithmic order of the number of qubits, achieving an exponential reduction of the required qubits compared to alternative methods, such as angle encoding. This method allows QFDNN to encode complex financial datasets with more features and fewer resources, supporting robust performance in tasks such as credit card fraud detection and loan prediction.
%This encoding strategy utilizes a combination of single-qubit $U$ gates and controlled operations to effectively encode classical information into quantum states across the multiple-qubit circuit. The above process is detailed in Algorithm \ref{alg:amplitude_encoding}. Figures \ref{Encode} and \ref{centroid} are the diagrammatic representations of the multiple qubit $UU^\dagger$ method, wherein the first case, the feature values of the data points are encoded sequentially (for $U$ circuit), whereas in the latter case, the feature values of the data points are encoded in reverse order (for $U^{\dagger}$ circuit). This encoding method uses a logarithmic order of number of qubits, which is an exponential reduction as compared to other methods such as angle encoding. Hence, with this encoding method, more features can be encoded using a lesser number of qubits.

%%%replacement for algorithm below 

\begin{algorithm}
\DontPrintSemicolon
\caption{Multiple Qubit Amplitude Encoding}
\label{alg:amplitude_encoding}

\SetKwFunction{ApplyGate}{apply\_gate}
\SetKwProg{Fn}{def}{:}{end}
\Fn{\ApplyGate{$qc, n$}}{
    \For{$i \leftarrow 1$ \KwTo $n$}{
        \If{$i = 1$}{
            Apply a single $U$ gate to $q_1$\;
        }
        \ElseIf{$i = 2$}{
            Apply a controlled $U$ gate on $q_2$ with $q_1$ as controller\;
            Apply an anti-controlled $U$ gate on $q_2$ with $q_1$ as anti-controller\;
        }
        \ElseIf{$i = 3$}{
            Apply a $U$ gate on $q_3$ with both $q_1$ and $q_2$ as controllers\;
            Apply a $U$ gate on $q_3$ with $q_1$ as controller and $q_2$ as anti-controller\;
            Apply a $U$ gate on $q_3$ with $q_1$ as anti-controller and $q_2$ as controller\;
            Apply a $U$ gate on $q_3$ with $q_1$ as anti-controller and $q_2$ as anti-controller\;
        }
        \ElseIf{$i = 4$}{
            \For{\text{each control/anti-control combination of } $q_1$, $q_2$, \text{and } $q_3$}{
                Apply a $U$ gate to $q_4$\;
            }
        }
        \Else{
            Repeat the process for higher qubits\;
        }
    }
    \textbf{Output:} Quantum circuit $qc$ with encoded information\;
}
\label{amp}
\end{algorithm}

%%% changed this 

% \begin{algorithm}
% \DontPrintSemicolon
% \caption{Multiple Qubit Amplitude Encoding}
% \label{alg:amplitude_encoding}

% \SetKwFunction{ApplyGate}{apply\_gate}
% \SetKwProg{Fn}{def}{:}{end}
% \Fn{\ApplyGate{$qc, n$}}{
%     \For{$i \leftarrow 1$ \KwTo $n$}{
%         \If{$i = 1$}{
%             \nonl \text{Apply a single } $U$ \text{ gate to } $q_1$\;
%         }
%         \ElseIf{$i = 2$}{
%             \nonl \text{Apply a controlled } $U$ \text{ gate on } $q_2$ \text{ with } $q_1$ \text{ as controller}\;
%             \newline
%             \nonl \text{Apply an anti-controlled } $U$ \text{ gate on } $q_2$ \text{ with } $q_1$ \text{ as anti-controller}\;
%         }
%         \ElseIf{$i = 3$}{
%             \nonl \text{Apply a } $U$ \text{ gate on } $q_3$ \text{ with both } $q_1$ \text{ and } $q_2$ \text{ as controllers}\;
%              \newline
%             \nonl \text{Apply a } $U$ \text{ gate on } $q_3$ \text{ with } $q_1$ \text{ as controller and } $q_2$ \text{ as anti-controller}\;
%              \newline
%             \nonl \text{Apply a } $U$ \text{ gate on } $q_3$ \text{ with } $q_1$ \text{ as anti-controller and } $q_2$ \text{ as controller}\;
%         }
%         \ElseIf{$i = 4$}{
%     \For{\text{each control/anti-control combination of } $q_1$, $q_2$, \text{and } $q_3$}{
%         Apply a $U$ gate to $q_4$\;
%     }
% }

%         \Else{
%             \nonl \text{Repeat the process for higher qubits}\;
%         }
%     }
%     \nonl \textbf{Output:} \text{Quantum circuit } $qc$ \text{ with encoded information}\;
% }
% \end{algorithm}

\subsection{Variational Quantum Circuit}
The proposed quantum circuit employs a parameterized approach to optimize performance in financial decision-making tasks by capturing the complexity of feature interactions while ensuring high accuracy (see Algorithm \ref{alg:parametrized_circuit}). The circuit begins with a parameterized Ry gate layer followed by an RZ gate layer. A series of CNOT gates is applied to entangle the qubits in the next layer. This is followed by a second layer of parameterized RY rotation gates to complete the variational structure, as shown in Fig. \ref{variational}. The execution of the circuit follows a variational approach in which the parameters are iteratively optimized using the COBYLA optimizer, a derivative-free classical optimization method. It calculates the cost function according to the initial parameter value. After that, the parameters are modified on the basis of the cost function, and a new output is evaluated. The process continues until the optimal parameter set is achieved, resulting in the best possible accuracy. This process enables QFDNN to improve predictive and decision-making capabilities. The loss function for the QFDNN architecture is defined as:

\begin{eqnarray}
\text{Loss} &=& 1 - \text{Accuracy}
\end{eqnarray}

The pseudocode for the overall model is provided in Algorithm \ref{alg:QFDNN}
\begin{algorithm}
\DontPrintSemicolon
\caption{Parametrized Circuit Construction}
\label{alg:parametrized_circuit}

\SetKwFunction{ConstructCircuit}{construct\_circuit}
\SetKwProg{Fn}{def}{:}{end}
\Fn{\ConstructCircuit{$qc, params$}}{
    $n \gets qc.num\_qubits$\; \tcp{Number of qubits in the circuit}
    $c \gets 0$\; \tcp{Counter to track the position in the parameter list}

    \For{$i \leftarrow 0$ \KwTo $n-1$}{
        Apply $R_y(params[c])$ to qubit $i$\;
        $c \gets c + 1$\;
    }

    \For{$i \leftarrow 0$ \KwTo $n-1$}{
        Apply $R_z(params[c])$ to qubit $i$\;
        $c \gets c + 1$\;
    }

    \For{$i \leftarrow 0$ \KwTo $n-2$}{
        Apply a controlled-X ($CX$) gate from qubit $i$ to $i+1$\;
    }

    \For{$i \leftarrow 0$ \KwTo $n-1$}{
        Apply $R_y(params[c])$ to qubit $i$\;
        $c \gets c + 1$\;
    }

    \textbf{Output:} Quantum circuit $qc$
}
\end{algorithm}

\begin{algorithm}
\DontPrintSemicolon
\caption{Quantum Feature Deep Neural Network (QFDNN)}
\label{alg:QFDNN}

\SetKwFunction{PreprocessData}{PreprocessData}
\SetKwFunction{EncodeFeatures}{EncodeFeatures}
\SetKwFunction{VQCInitialization}{VQCInitialization}
\SetKwFunction{HybridTraining}{HybridTraining}
\SetKwFunction{PredictModel}{PredictModel}

\SetKwProg{Fn}{def}{:}{end}

\Fn{\PreprocessData{$D$}}{
    Normalize and preprocess dataset $D$\; 
    Apply Principal Component Analysis (PCA) for feature reduction\;  
    \textbf{Output:} Processed dataset $D'$
}

\Fn{\EncodeFeatures{$D', n$}}{
    $qreg \gets$ Quantum register with $n$ qubits\;  
    \For{each data sample $x_i \in D'$}{
        Encode features into quantum states using multiple-qubit UU$^\dagger$ method\;  
        Apply unitary transformations $U$ and $U^{\dagger}$ to encode data\;  
    }
    \textbf{Output:} Quantum-encoded feature set  
}

\Fn{\VQCInitialization{$U(\theta)$}}{
    Define parameterized quantum circuit $U(\theta)$\;  
    Initialize quantum parameters $\theta_s$\;  
    \textbf{Output:} Initialized variational quantum circuit
}

\Fn{\HybridTraining{$U(\theta), D', \text{Optimizer}$}}{
    \While{Loss not converged}{
        \For{each training sample $x_i$}{
            Apply $U(\theta)$ to quantum register after $U$ and $U^{\dagger}$\;  
            Measure quantum state to extract features\;  
            Compute loss function using a quantum-classical hybrid approach\;  
            Update $\theta$ using COBYLA optimizer\;  
        }
    }
    \textbf{Output:} Optimized quantum circuit parameters $\theta^*$
}

\Fn{\PredictModel{$U(\theta^*)$}}{
    Use trained $U(\theta^*)$ for classification\;  
    Compute accuracy, precision, recall, and F1-score\;  
    \textbf{Output:} Classification results  
}

\end{algorithm}

\subsection{Noise Models}
In QC, noise refers to environmental disturbances affecting quantum states. Kraus operators \(\{E_i\}\) model this noise by mapping quantum states while ensuring the completeness condition \(\sum_{i=0}^{N} E_{i}^{\dagger} E_{i} = I\), which preserves normalization and unit trace. The evolved state \(\rho'\) is given by 
\begin{equation}
\rho' = \sum_{i=0}^{N} E_{i} \rho E_{i}^{\dagger},
\end{equation} 
representing the impact of each operator on the initial state and allowing us to describe noise and interactions in a quantum system.

\subsubsection{Bitflip}
It occurs when a qubit's state flips between 0 and 1 with a certain probability. The Kraus operators for this error are:
\begin{equation}
E_{0} = \sqrt{1 - p}I, E_{1} = \sqrt{p}X.
\end{equation}

\subsubsection{Phaseflip}
It changes the qubit's phase, flipping the state from \(\ket{0} + \ket{1}\) to \(\ket{0} - \ket{1}\) or vice versa. The Kraus operators for this error are:
\begin{equation}
E_{0} = \sqrt{1 - p} I, E_{1} = \sqrt{p} Z.
\end{equation}

\subsubsection{BitPhase Flip}
It alters both the bit value and the phase of a qubit, transforming \(\ket{0}\) to \(-\ket{1}\) or vice versa. The Kraus operators for this error are:
\begin{equation}
E_{0} = \sqrt{1 - p}I, E_{1} = \sqrt{p} Y.
\end{equation}

\subsubsection{Depolarizing}
It is a stochastic quantum error in which a qubit can randomly rotate about any axis on the Bloch sphere with a certain probability, reflecting the strength of the noise. It describes the effect of an error channel that transforms the state of a qubit as \(\rho \longrightarrow (1 - p) \rho + \frac{p}{2} I\).
The Kraus operators for the depolarizing map can be expressed as:
\begin{eqnarray}
E_{0} = \sqrt{1 - \frac{3p}{4}}I, E_{1} = \sqrt{\frac{p}{4}} Z, E_{2} = \sqrt{\frac{p}{4}}X, E_{3} = \sqrt{\frac{p}{4}} Y.
\end{eqnarray}

\subsubsection{Amplitude Damping}
It is a common error in quantum systems due to energy dissipation. The Kraus operators for this error are:
\begin{equation}
E_{0} = \sqrt{p} \ket{0}\bra{1}, \quad E_{1} = \begin{bmatrix} 1 & 0 \\ 0 & \sqrt{1 - p} \end{bmatrix}.
\end{equation}

\subsubsection{Phase Damping}
It causes the loss of phase information, affecting the performance of the quantum algorithm. The Kraus operators for this error are:
\begin{equation}
E_{0} = \sqrt{1 - p}I, E_{1} = \sqrt{p} \ket{0}\bra{0}, E_{2} = \sqrt{p} \ket{1}\bra{1}.
\end{equation}

\section{Experimental Results \label{SecIV}}

\subsection{Hyperparameters\label{hyperparameters}}
The model parameters are optimized using the COBYLA algorithm. Training is performed over 10 iterations, which provides stable convergence for both datasets. The circuit's parameters are initialized randomly, and the optimization process employs a complement of an accuracy loss function to guide learning. During training, parameterized RY and RZ gate rotations are updated iteratively to enhance the model's performance. Performance is analyzed by plotting the accuracy against noise strength for both datasets, which provides insight into how the QFDNN handles real-world quantum computing conditions.
%The model's parameters are optimized using the COBYLA algorithm. Training is performed over 10 iterations, which provides stable convergence for both the datasets. The circuit's parameters are initialized randomly, and the optimization process employs a complement of an accuracy loss function to guide learning. Parameterized RY and RZ gate rotations are updated iteratively during training to enhance the model's performance. Performance is analyzed by plotting the accuracy against noise strength for both datasets, which provides insights into how the QFDNN handles real-world quantum computing conditions.

\subsection{Datasets}\label{QVP:Sec2}
The datasets considered for this QFDNN are the Credit Card Fraud (CCF) dataset \cite{beach2024creditcardfraud} and the Loan Prediction (LP) dataset \cite{zohaib2024eligibilityprediction}. The CCF dataset consists of transaction data from European cardholders, with the goal of identifying fraudulent transactions. A balanced dataset is created with an equal number of fraudulent and non-fraudulent transactions. The dataset originally contained 30 features, of which the features ``Amount" and ``Time" are excluded. Then, PCA is applied to reduce the number of features from 28 to 7. The LP dataset is provided by Dream Housing Finance Company and is used to automate the loan eligibility process. With 11 features in total, PCA is applied similarly to reduce the number of features to 7 for simplicity.

\subsection{Preprocessing}
In preprocessing for the CCF and LP datasets, the minority sample is increased using the Synthetic Minority Oversampling Technique (SMOTE). The sample size for both CCF and LP is 500 cases, discarding others for compatibility with the simulator. Data is trained and tested on the same 500 data points for classical and QFDNN for a fair comparison. After scaling, PCA reduces the original 28 features to 7 principal components before feeding them to QFDNN. Similarly, the LP dataset, which initially had 11 features in number, follows a similar preprocessing strategy. Data is scaled and then reduced via PCA before being input into the models. 
%In preprocessing the CCF dataset, it is first balanced to ensure equal representation of fraud and non-fraud instances. The sample size is shrunk to 100 instances for compatibility with the simulator. For a fair comparison, data is trained and tested on the same 100 data points for both classical and QFDNN. After scaling, PCA is used to reduce the original 28 features to 7 principal components before feeding to QFDNN. Similarly, the LP dataset, which originally had 11 features in number, follows a similar preprocessing strategy. Data is scaled and then reduced via PCA before its input into the models. However, unlike the CCF dataset, the LP dataset is not balanced. The first 100 data points are used for LP rather than balancing the data for classical and quantum. This allows both datasets to fit their respective models and preserve key features in this dimensionality reduction while also maintaining a fair comparison.

\subsection{Comparative Models and Evaluation Metrics}
Several well-established ML algorithms are implemented to provide a comparative benchmark for the proposed QFDNN model. Each algorithm is configured with specific hyperparameters and settings to optimize performance. Logistic Regression (LR) is implemented with L2 regularization and the `lbfgs' solver. The regularization strength is set to the default value of C=1.0. The K-Nearest Neighbors (KNN) classifier is applied with the default setting of 5 neighbours, where all neighbours are weighted equally, and the algorithm is set to `auto,' allowing the model to choose the best algorithm based on the dataset. A support vector machine (SVM) is utilized with the RBF kernel, selected for its ability to handle non-linear classification tasks. The C parameter is set to 1.0, controlling the margin of separation, and the gamma is set to 'scale', which automatically adjusts based on the number of features in the data. For the artificial neural network (ANN), the model is trained with a maximum of 1000 iterations and the default ReLU activation function, using the `Adam' optimizer, which is efficient for large datasets. The random forest (RF) classifier employs 100 estimators (trees), with the default Gini impurity criterion for splitting nodes and no limit on the depth of the trees, which allows the trees to grow until all leaves are pure or contain fewer than a specified number of samples. Finally, gradient boosting (GB) is configured with 100 estimators, a learning rate of 0.1, and a maximum depth of 3 to mitigate overfitting while ensuring balanced performance. The performance of models is evaluated using metrics such as accuracy, precision, recall, and the F1 score on the CCF and LP datasets. 

\subsection{Results of CCF}

\begin{figure*}[ht!]
\centering
\begin{subfigure}{.49\linewidth}
    \centering
    \includegraphics[width=\linewidth]{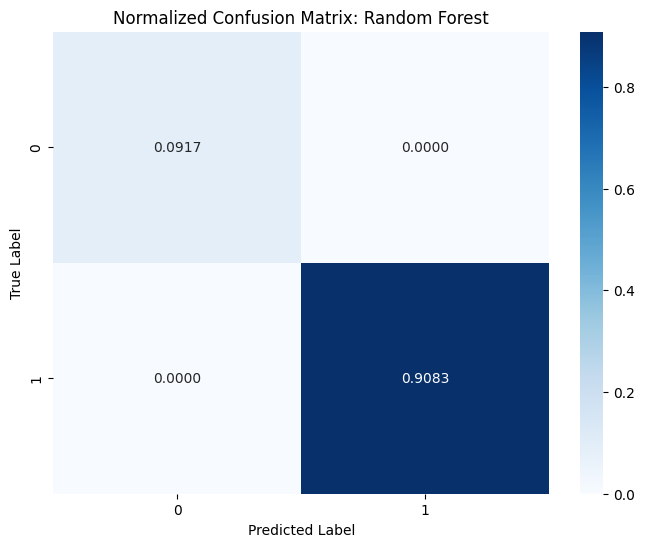 }
    \caption{}
    \label{subfig1classical}
\end{subfigure}
\hfill
\begin{subfigure}{.49\linewidth}
    \centering
    \includegraphics[width=\linewidth]{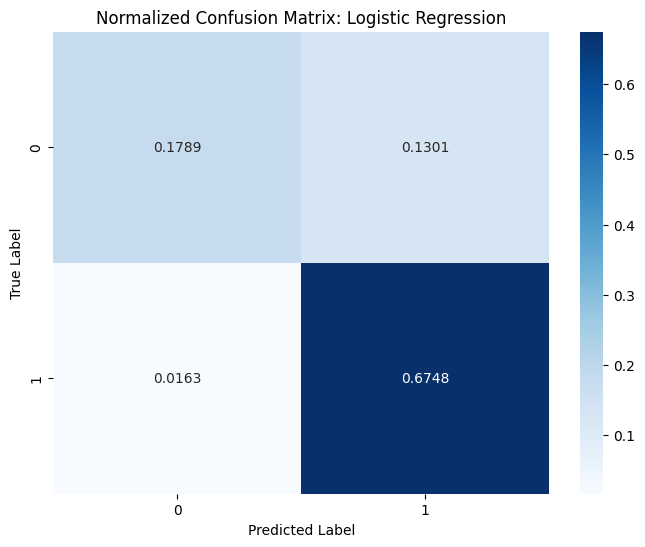}
    \caption{}
    \label{subfig2classical}
\end{subfigure}\hfill
\begin{subfigure}{.49\linewidth}
    \centering
    \includegraphics[width=\linewidth]{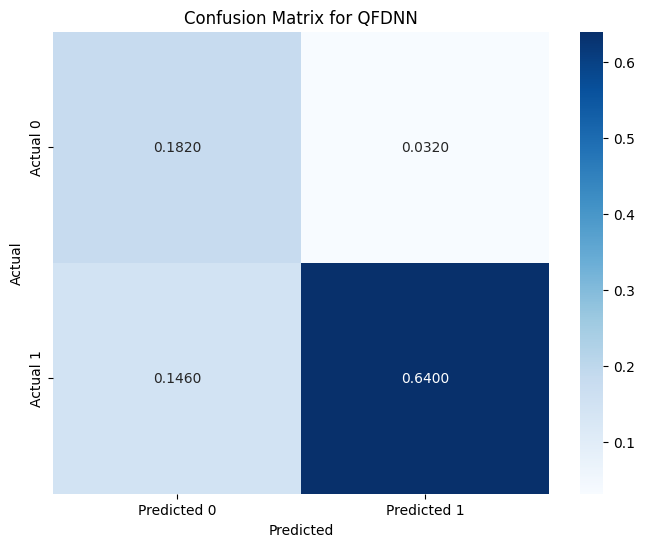}
    \caption{}
    \label{subfig1quantum}
\end{subfigure}
\hfill
\begin{subfigure}{.49\linewidth}
    \centering
    \includegraphics[width=\linewidth]{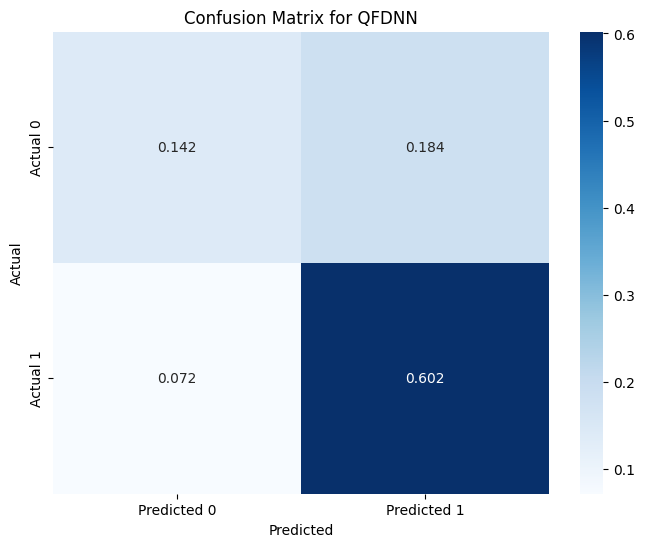}
    \caption{}
    \label{subfig2quantum}
\end{subfigure}
\caption{Confusion Matrices of LR: (a) CCF dataset and (b) LP dataset. Confusion Matrices of QFDNN: (c) CCF dataset and (d) LP dataset.}
\label{fig:Confusion_matrixclassical}
\end{figure*}

\begin{figure*}[]
\centering
\begin{subfigure}{.5\linewidth}
    \includegraphics[width=\linewidth]{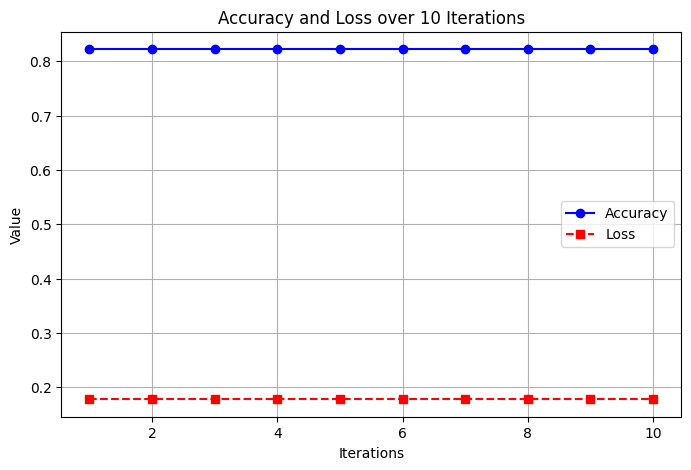}
    \caption{}
    \label{subfig1b}
\end{subfigure}\hfill
\begin{subfigure}{.5\linewidth}
    \includegraphics[width=\linewidth]{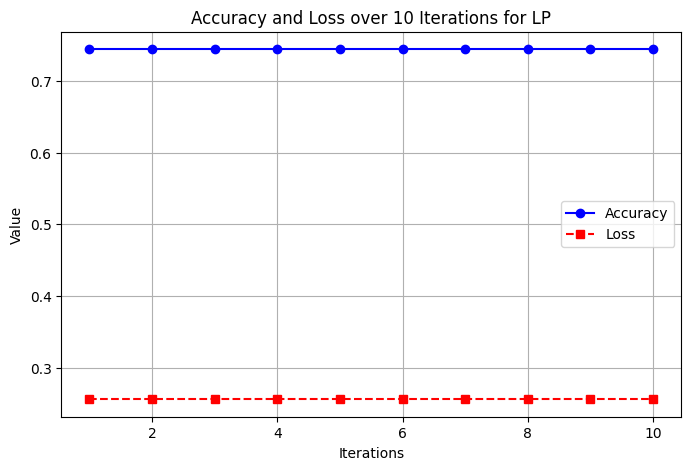 }
    \caption{}
    \label{subfig1a}
\end{subfigure}\hfill
\caption{\normalsize{Accuracy and Loss versus Iterations for (a) CCF dataset and (b) LP dataset.}}
\label{Accuracy}
\end{figure*}

The performance of the CCF dataset is evaluated against various classical ML algorithms, as shown in Table \ref{CCF_comp}. The QFDNN model is applied to the same dataset for the quantum approach. Initially, the dataset consisted of 28 features and was reduced to seven features using PCA to simplify the input space while retaining critical information. The QFDNN model is trained over 10 iterations with the COBYLA optimization method, which iteratively adjusts the variational parameters to minimize the loss function and enhance accuracy. The QFDNN has been shown to achieve a maximum accuracy of 0.822 with constant learning in all iterations, with a steady increase in accuracy and a corresponding decrease in loss before stable at 0.822. The QFDNN loss stabilizes at 0.178, indicating that the model approaches but does not fully converge to an optimal solution, as shown in Fig. \ref{Accuracy}. 
The confusion matrix for the CCF dataset offers valuable information on the performance of the models. It highlights the QFDNN's effectiveness in correctly classifying negative instances while also revealing its limitations in identifying all true positive cases, as shown in Fig. \ref{fig:Confusion_matrixclassical}. The performance metrics of classical ML demonstrated that LR, SVM, ANN, RF, KNN, and GB achieve nearly perfect scores of 1.000 across all metrics (see Table \ref{CCF_comp}). The best performance among the classical models is achieved by RF, prompting the creation of a confusion matrix to represent its results (Fig. \ref{subfig1classical}). However, despite the high performance of the classical models, QFDNN demonstrates unique strengths, particularly achieving a near-perfect precision score of 0.9524. The QFDNN achieves an F1 score of 0.8779, reflecting a trade-off, with a recall of 0.8142, highlighting areas for improvement in identifying true positives (Fig. \ref{subfig1quantum}). In addition, to understand the effectiveness of each component, an ablation study was performed for the variational and $UU^\dagger$ components. $UU^\dagger$ demonstrates a strong accuracy of 0.726, and for the variational part, the accuracy was 0.786. Tables \ref{tab:t_test_ccf} and \ref{tab:t_test_lp} represent a t-test and p-value study between different classical models and QFDNN for the CCF and LP datasets, respectively.

\begin{table}[]
    \centering
    \begin{tabular}{|c|c|c|c|c|}
    \hline 
      \textbf{Model} & \textbf{Accuracy} & \textbf{Precision} & \textbf{Recall} & \textbf{F1 Score} \\
    \hline
          LR & 0.9908 & 1.000 & 0.9899 & 0.9949 \\
         \hline
         KNN & 0.9633 & 1.000 & 0.9596 & 0.9794 \\
         \hline
     SVM & 0.9908 & 0.990 & 1.000 & 0.9950 \\
         \hline
      ANN & 0.9908 & 0.990 & 1.000 & 0.9950 \\
         \hline
    RF & 1.0000 & 1.000 & 1.000 & 1.000 \\
         \hline
    GB & 1.0000 & 1.000 & 1.000 & 1.000 \\
         \hline
    \textbf{QFDNN} & \textbf{0.822} & \textbf{0.9524} & \textbf{0.8779} & \textbf{0.814} \\
         \hline
    \end{tabular}
    \caption{Comparative Evaluation of QFDNN Against Classical Models for CCF}
    \label{CCF_comp}
\end{table}

\begin{figure*}[]
\centering
\begin{subfigure}{.5\linewidth}
    \includegraphics[width=\linewidth]{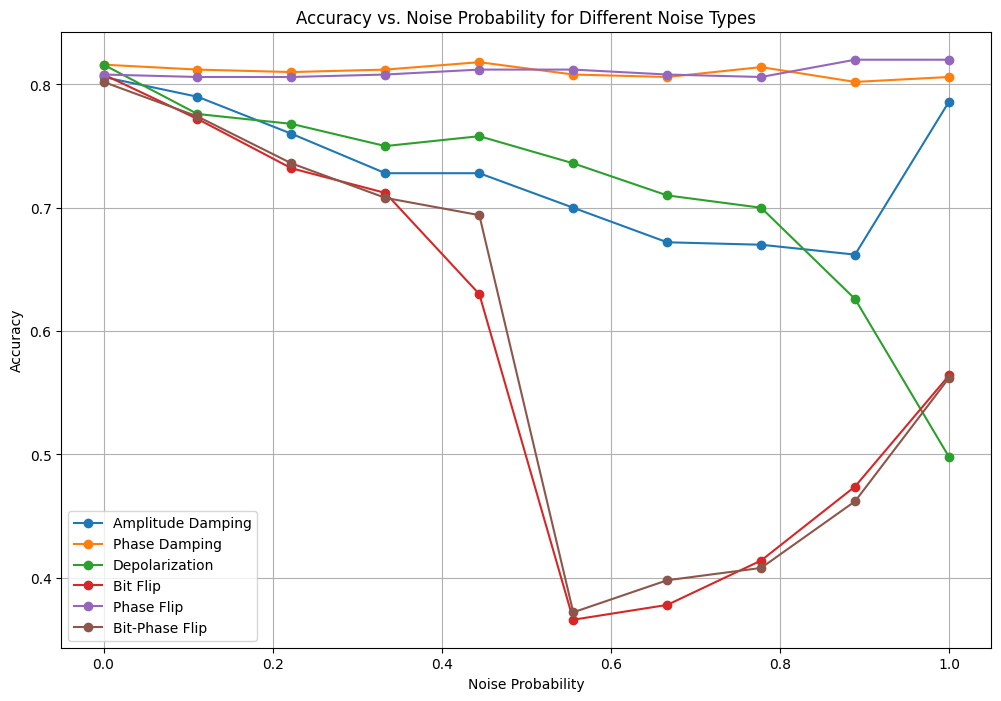}
    \caption{}
    \label{subfig1b}
    
\end{subfigure}\hfill
\begin{subfigure}{.5\linewidth}
    \includegraphics[width=\linewidth]{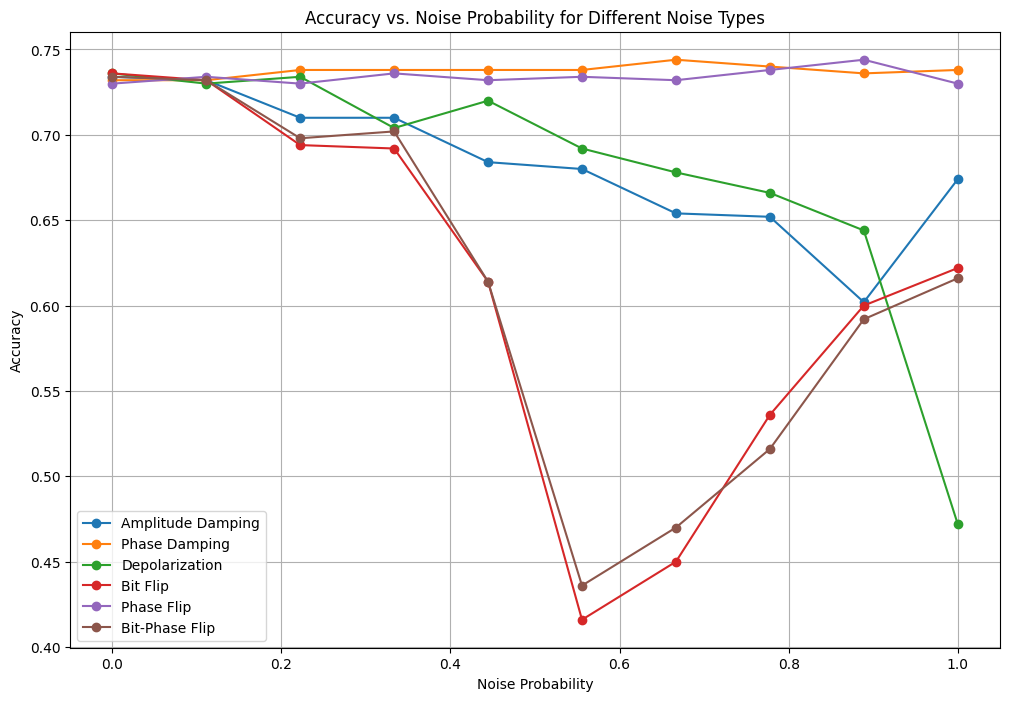 }
    \caption{}
    \label{subfig1a}
\end{subfigure}\hfill

\caption{\normalsize{Accuracy versus Noise Strength for Six Types of Noise for (a) CCF dataset and (b) LP dataset.}}
\label{noiseplots}
\end{figure*}

\subsection{Results of LP} 

Table \ref{L_comp} provides a detailed comparison of the classical models against QFDNN for the LP dataset. The QFDNN model is applied to the LP dataset, which initially includes 11 features and is reduced to 7 using PCA, simplifying the input space while retaining critical information. The QFDNN model undergoes 10 iterations of training with the COBYLA optimization method. The QFDNN achieves a peak accuracy of 0.744, which plateaus after successive iterations, with the loss stabilizing at 0.256, as shown in Fig. \ref{Accuracy}. The performance of the classical models demonstrated that LR stands out in almost all metrics. LR follows closely, achieving an accuracy of 0.854 and an F1 score of 0.902. 

The QFDNN model shows competitive performance, achieving an accuracy of 0.744, which is close to the accuracy of the ANN (0.772). The QFDNN precision is 0.7659. The recall of QFDNN at 0.8932 surpasses the ANN classical models and is close to other classical models. The QFDNN's F1 score of 0.8247 highlights its robust balance between precision and recall. The confusion matrix for the LP dataset highlights the model's classification capability. This analysis underscores the QFDNN model's ability to deliver competitive performance in predictive analytics for the LP dataset. While LR and SVM achieve higher overall metrics, the QFDNN's performance demonstrates its potential, particularly in scenarios where recall and feature reduction are critical considerations. The confusion matrices for the best classical case (LR) and QFDNN are shown in Figures \ref{subfig2classical} and \ref{subfig2quantum}, respectively. Similarly, to understand the effectiveness of each component, an ablation study was conducted for the variational and $UU^\dagger$ components. $UU^\dagger$ demonstrates an accuracy of 0.634, and for the variational part, the accuracy was 0.674.
\begin{table}
\centering
\begin{tabular}{|c|c|c|c|c|}
\hline
\textbf{Algorithm} & \textbf{Accuracy} & \textbf{Precision} & \textbf{Recall} & \textbf{F1 Score} \\
\hline
Logistic Regression & 0.854 & 0.838 & 0.976 & 0.902 \\
\hline
KNN & 0.846 & 0.837 & 0.965 & 0.896 \\
\hline
SVM & 0.854 & 0.832 & 0.988 & 0.903 \\
\hline
ANN & 0.772 & 0.813 & 0.871 & 0.841 \\
\hline
Random Forest & 0.821 & 0.839 & 0.918 & 0.876 \\
\hline
Gradient Boosting & 0.813 & 0.823 & 0.929 & 0.873 \\
\hline
\textbf{QFDNN} & \textbf{0.744} & \textbf{0.7659} & \textbf{0.8932} & \textbf{0.8247} \\
\hline
\end{tabular}
\caption{Comparative Evaluation of QFDNN Against Classical Models for LP.}
\label{L_comp}
\end{table}

\begin{table}
\centering
\begin{tabular}{|c|c|c|c|}
\hline
\textbf{Model} & \textbf{T-Statistic} & \textbf{P-Value} & \textbf{Significant Difference?} \\
\hline
LR & $-\infty$ & 0.0000 & \textbf{Yes} \\
\hline
KNN & $-\infty$ & 0.0000 & \textbf{Yes} \\
\hline
SVM & -12.5757 & 0.0002 & \textbf{Yes} \\
\hline
ANN & -23.2525 & 0.0000 & \textbf{Yes} \\
\hline
RF & $-\infty$ & 0.0000 & \textbf{Yes} \\
\hline
GB & $-\infty$ & 0.0000 & \textbf{Yes} \\
\hline
\end{tabular}
\caption{T-Test and P-Value  Comparing QFDNN Against Other Models for CCF Dataset.}
\label{tab:t_test_ccf}
\end{table}

\begin{table}
\centering
\begin{tabular}{|c|c|c|c|}
\hline
\textbf{Model} & \textbf{T-Statistic} & \textbf{P-Value} & \textbf{Significant Difference?} \\
\hline
LR & -3.5535 & 0.0237 & \textbf{Yes} \\
\hline
KNN & 5.8630 & 0.0042 & \textbf{Yes} \\
\hline
SVM & 1.8899 & 0.1318 & No \\
\hline
ANN & 2.3292 & 0.0803 & No \\
\hline
RF & -2.1157 & 0.1018 & No \\
\hline
GB & -2.9842 & 0.0406 & \textbf{Yes} \\
\hline
\end{tabular}
\caption{T-Test and P-Value  Comparing QFDNN Against Other Models for LP Dataset.}
\label{tab:t_test_lp}
\end{table}

\subsection{Effect of Noise on the Model}
Assessing the model's resilience to various quantum noise types is critical for practical quantum computing applications. The impact of various noisy environments on the QFDNN accuracy for the CCF and LP datasets is shown in Fig. \ref{noiseplots}.

For the CCF dataset, the QFDNN achieves its highest accuracy of 0.822 without noise, serving as the baseline. The accuracy drops significantly to 0.366 and 0.372 for bit-flip and bit-phase flip noise at noise strength 0.55 for both noise types, respectively, then increases as the noise strength increases, highlighting the model's vulnerability to changes in bit states and combined bit-phase states. In contrast, phase flip noise demonstrates remarkable stability, maintaining a near-constant accuracy at 0.822 across all noise levels, suggesting strong resilience to pure phase disruptions. Amplitude damping noise causes a moderate decline in accuracy, gradually reducing from 0.830 to 0.662, then increasing afterwards. The depolarization noise leads to an accuracy dropping from 0.830 to 0.498. Finally, phase damping noise shows minimal impact, with accuracy remaining steady at approximately 0.822, demonstrating robustness to phase coherence loss.

For the LP dataset, QFDNN achieves its highest accuracy of 0.744 without noise. The accuracy drops significantly to 0.416 and 0.436 for bit-flip and bit-phase flip noise as the noise strength reaches 0.55, emphasizing the model's sensitivity to bit-related disruptions. Phase flip noise remains stable, with accuracy fluctuating slightly around 0.734, suggesting minimal impact. Amplitude damping noise gradually decreases accuracy from 0.734 to 0.602 with noise strength 0.88, increasing afterwards. Meanwhile, depolarization noise causes a more pronounced decrease of 0.472. Phase damping noise has the least impact, with an accuracy of around 0.734 throughout the noise strength range.
Across both datasets, the QFDNN model's performance is most affected by bit-flip and bit-phase flip noise, which cause substantial accuracy reductions as noise strength increases, reflecting the model's sensitivity to changes in qubit states. However, phase flip noise demonstrates a consistent trend of minimal impact with stable accuracy. Amplitude damping and depolarization noise exhibit intermediate effects, causing gradual declines in accuracy, whereas phase damping noise has the least impact, maintaining stable performance. 
These findings underline the necessity for error mitigation strategies, especially for bit flip and bit-phase flip noise, to enhance the QFDNN's robustness and applicability in noisy quantum environments for financial decision-making tasks.

\begin{table}[]
    \centering
    \begin{tabular}{|c|c|c|c|}
    \hline 
      \textbf{Model} &\textbf{Accuracy} & \textbf{Precision} & \textbf{Recall}\\
    \hline
          SVM \cite{RB202135} &0.934 &0.974 &0.897 \\
         \hline
         KNN  \cite{RB202135} & 0.998&0.714 & 0.039\\
         \hline
     ANN  \cite{RB202135} &0.999 & 0.811 &0.761\\
         \hline
      \textbf{QFDNN} & \textbf{0.822} & \textbf{0.9524} & \textbf{0.8779} \\
         \hline
    \end{tabular}
    \caption{Comparative Analysis of QFDNN with Existing Classical Models for CCF.}
    \label{tab:my_label}
\end{table}

\begin{table}[h!]
    \centering
    \begin{tabular}{|c|c|c|c|}
    \hline 
      \textbf{Model} & \textbf{Accuracy} & \textbf{Precision} & \textbf{Recall} \\
    \hline
      ABC \cite{haque2024bankloanpredictionusing}& 0.999 & 0.999 & 0.999 \\
      \hline
      RFC \cite{haque2024bankloanpredictionusing}& 0.999 & 0.999 & 0.999 \\
      \hline
      SVM \cite{haque2024bankloanpredictionusing}& 0.998 & 0.998 & 0.998 \\
      \hline
      DC \cite{haque2024bankloanpredictionusing}& 0.999 & 0.999 & 0.999 \\
      \hline
      GN \cite{haque2024bankloanpredictionusing}& 0.771 & 0.803 & 0.771 \\
      \hline
      \textbf{QFDNN} & \textbf{0.744} & \textbf{0.7659} & \textbf{0.8932} \\
      \hline
    \end{tabular}
    \caption{Comparative Analysis of QFDNN with Existing Classical Models for LP.}
    \label{tab:my_label2}
\end{table}

\begin{table}[h!]
    \centering
    \begin{tabular}{|l|c|c|c|c|}
        \hline
        \textbf{Algorithm} & \textbf{CCF} & \textbf{LP} & \textbf{Features} & \textbf{Qubits} \\
        \hline
        QFDNN  & \checkmark & \checkmark & $2^{N}-1$ & 3 \\
        \hline
        LEP QNN \cite{innan2024lepqnnloaneligibilityprediction} & \texttimes & \checkmark & $N$ & 7\\
        \hline
        FHQNN \cite{math12091391} & \checkmark & \texttimes & $N$ & 12 \\
        \hline
        QGNN \cite{Innan_2024} & \checkmark & \texttimes & $N$ & 7\\
        \hline
    \end{tabular}
    \caption{Comparative Analysis of Required Qubits to Encode Features between QFDNN with Existing Quantum Models for CCF and LP }
    \label{tab:quantum_algorithms}
\end{table}

\begin{table}[h!]
    \centering
    \begin{tabular}{|l|c|c|}
        \hline
        \textbf{Algorithm} & \textbf{Gate Complexity} & \textbf{Time} \\
        \hline
        QFDNN  &$O(2^N)$ & $O(2^N)$ \\
        \hline
        LEP QNN \cite{innan2024lepqnnloaneligibilityprediction} &$O(dN)$ & $O(dN)$\\
        \hline
        FHQNN \cite{math12091391} & $O(dN)$ & $O(N^{2})$ \\
        \hline
        QGNN \cite{Innan_2024} & $O(dN^2)$ &$O(d\log(N) + E + V\log(V))$ \\
        \hline
    \end{tabular}
    \caption{Gate Complexity and Time Analysis of QFDNN Compared to Existing Quantum Models, Where \(d\) Denotes the Depth of the Variational Component, and \(E\) and \(V\) Represent the Edges and Vertices of the Graph, Respectively.}
    \label{tab:quantum_algorithms_gate_time}
\end{table}

%FULL HYBRID (FH) Quantum Neural Network Model
\section{Discussion and Conclusion\label{SecV}}
This study introduces the Quantum Feature Deep Neural Network (QFDNN) as a resource-efficient and noise-resilient model for financial decision-making tasks, specifically credit card fraud detection and loan eligibility prediction. The QFDNN utilizes variational quantum circuits and the $UU^{\dagger}$ method for effective feature representation; it achieves maximum accuracies of 0.822 and 0.744 on the respective datasets while maintaining robustness against various noise models. The performance analysis of QFDNN for the qubits required to encode the features, gate complexity, and required time compared to various quantum models such as the Full HQNN Model (FHQNN) \cite{math12091391}, which addresses the CCF problem with 12 qubits and approximately 350 iterations, also with the LEP QNN \cite{innan2024lepqnnloaneligibilityprediction} and QGNN models which provide alternative approaches, focusing on graph-based representations for financial problem solving, is shown in Tables \ref{tab:quantum_algorithms} and \ref{tab:quantum_algorithms_gate_time}. QFDNN demonstrates resource efficiency and applicability with fewer qubits than other quantum models. Furthermore, QFDNN is compared against classical classifiers such as AdaBoost (ABC), DecisionTree (DTC), GaussianNB (GN), RandomForest (RFC), and SVM \cite{haque2024bankloanpredictionusing} (see Tables \ref{tab:my_label} and \ref{tab:my_label2}). It highlights the potential of QFDNN to handle complex, high-dimensional data and serves as a robust solution for automating financial analytics.
The QFDNN's sensitivity to bit flip and bit-phase flip noise highlights the necessity for error mitigation strategies to enhance the QFDNN's robustness and applicability in noisy quantum environments for financial decision-making tasks. A possible solution would be to integrate quantum noise mitigation techniques such as zero noise extrapolation (ZNE) \cite{9259940}. Furthermore, while QFDNN offers potential advantages in processing multi-dimensional data, its performance is susceptible to the choice of optimization methods and preprocessing quality. The QFDNN used the COBYLA optimizer, which ensures stable convergence, but the training process takes comparatively more time. Our choice of the COBYLA optimizer was driven by the nature of our cost function. Since this function is non-differentiable in practice, gradient-based optimizers like Adam often fail to provide meaningful updates, frequently stagnating around 50\% accuracy. COBYLA, being a derivative-free optimizer, efficiently explores the parameter space using linear approximations, ensuring stable convergence. Therefore, future work can focus on exploring alternative optimizers such as Adam or Gradient Descent with different cost functions, and using deeper variational circuits could further enhance accuracy and efficiency. Finally, QFDNN represents a promising approach to quantum-enhanced financial analytics, particularly in scenarios with complex and reduced feature sets. The practical deployment of QFDNNs is expected to become increasingly feasible with the advancement of quantum computing hardware, offering efficient, scalable, and robust solutions for critical financial decision-making tasks. Future research on QFDNN can expand its impact by adapting it to broader financial tasks like portfolio optimization and risk assessment, and by implementing hardware-specific optimizations for emerging quantum processors. To improve interpretability and transparency, techniques such as Shapley values, quantum Fisher information, and post-hoc explainability methods such as SHAP and LIME can be applied to analyze feature contributions and support decision justification in financial applications.

\ifCLASSOPTIONcaptionsoff
\newpage
\fi

\bibliographystyle{IEEEtran}

\bibliography{IEEE}

\begin{thebibliography}{10}
\providecommand{\url}[1]{#1}
\csname url@samestyle\endcsname
\providecommand{\newblock}{\relax}
\providecommand{\bibinfo}[2]{#2}
\providecommand{\BIBentrySTDinterwordspacing}{\spaceskip=0pt\relax}
\providecommand{\BIBentryALTinterwordstretchfactor}{4}
\providecommand{\BIBentryALTinterwordspacing}{\spaceskip=\fontdimen2\font plus
\BIBentryALTinterwordstretchfactor\fontdimen3\font minus \fontdimen4\font\relax}
\providecommand{\BIBforeignlanguage}[2]{{%
\expandafter\ifx\csname l@#1\endcsname\relax
\typeout{** WARNING: IEEEtran.bst: No hyphenation pattern has been}%
\typeout{** loaded for the language `#1'. Using the pattern for}%
\typeout{** the default language instead.}%
\else
\language=\csname l@#1\endcsname
\fi
#2}}
\providecommand{\BIBdecl}{\relax}
\BIBdecl

\bibitem{8428484}
L.~Zheng, G.~Liu, C.~Yan, and C.~Jiang, ``Transaction fraud detection based on total order relation and behavior diversity,'' \emph{IEEE Transactions on Computational Social Systems}, vol.~5, no.~3, pp. 796--806, 2018.

\bibitem{ebiaredoh2021artificial}
S.~A. Ebiaredoh-Mienye, E.~Esenogho, and T.~G. Swart, ``Artificial neural network technique for improving prediction of credit card default: A stacked sparse autoencoder approach,'' \emph{International Journal of Electrical and Computer Engineering}, vol.~11, no.~5, p. 4392, 2021.

\bibitem{Hawley1990}
\BIBentryALTinterwordspacing
D.~D. Hawley, J.~D. Johnson, and D.~Raina, ``Artificial neural systems: A new tool for financial decision-making,'' \emph{Financial Analysts Journal}, vol.~46, no.~6, pp. 63--72, 1990. [Online]. Available: \url{https://doi.org/10.2469/faj.v46.n6.63}
\BIBentrySTDinterwordspacing

\bibitem{10491318}
P.~Xia, X.~Zhu, V.~Charles, X.~Zhao, and M.~Peng, ``A novel heuristic-based selective ensemble prediction method for digital financial fraud risk,'' \emph{IEEE Transactions on Engineering Management}, vol.~71, pp. 8002--8018, 2024.

\bibitem{10242155}
X.~Cao and S.~Li, ``Neural networks for portfolio analysis with cardinality constraints,'' \emph{IEEE Transactions on Neural Networks and Learning Systems}, vol.~35, no.~12, pp. 17\,674--17\,687, 2024.

\bibitem{10613029}
K.~Alam, M.~H. Bhuiyan, I.~U. Haque, M.~F. Monir, and T.~Ahmed, ``Enhancing stock market prediction: A robust lstm-dnn model analysis on 26 real-life datasets,'' \emph{IEEE Access}, vol.~12, pp. 122\,757--122\,768, 2024.

\bibitem{TAN2024102049}
\BIBentryALTinterwordspacing
J.~Tan, M.~Deveci, J.~Li, and K.~Zhong, ``Asset pricing via fused deep learning with visual clues,'' \emph{Information Fusion}, vol. 102, p. 102049, 2024. [Online]. Available: \url{https://www.sciencedirect.com/science/article/pii/S1566253523003652}
\BIBentrySTDinterwordspacing

\bibitem{KWON2025125327}
\BIBentryALTinterwordspacing
S.~Kwon, J.~Jang, and C.~O. Kim, ``Credit scoring using multi-task siamese neural network for improving prediction performance and stability,'' \emph{Expert Systems with Applications}, vol. 259, p. 125327, 2025. [Online]. Available: \url{https://www.sciencedirect.com/science/article/pii/S0957417424021948}
\BIBentrySTDinterwordspacing

\bibitem{GRUDNIEWICZ2023102052}
\BIBentryALTinterwordspacing
J.~Grudniewicz and R.~Ślepaczuk, ``Application of machine learning in algorithmic investment strategies on global stock markets,'' \emph{Research in International Business and Finance}, vol.~66, p. 102052, 2023. [Online]. Available: \url{https://www.sciencedirect.com/science/article/pii/S0275531923001782}
\BIBentrySTDinterwordspacing

\bibitem{culkin2017machine}
\BIBentryALTinterwordspacing
R.~Culkin and S.~R. Das, ``Machine learning in finance: the case of deep learning for option pricing,'' \emph{Journal of Investment Management}, vol.~15, no.~4, pp. 92--100, 2017. [Online]. Available: \url{https://srdas.github.io/Papers/BlackScholesNN.pdf}
\BIBentrySTDinterwordspacing

\bibitem{SNASEL2024102018}
\BIBentryALTinterwordspacing
V.~Snášel, J.~D. Velásquez, M.~Pant, D.~Georgiou, and L.~Kong, ``A generalization of multi-source fusion-based framework to stock selection,'' \emph{Information Fusion}, vol. 102, p. 102018, 2024. [Online]. Available: \url{https://www.sciencedirect.com/science/article/pii/S1566253523003342}
\BIBentrySTDinterwordspacing

\bibitem{sen2022introductory}
J.~Sen, R.~Sen, and A.~Dutta, ``Introductory chapter: machine learning in finance-emerging trends and challenges,'' \emph{Algorithms, Models and Applications}, p.~1, 2022.

\bibitem{ARAUJO2023100087}
\BIBentryALTinterwordspacing
G.~S. Araujo and W.~P. Gaglianone, ``Machine learning methods for inflation forecasting in brazil: New contenders versus classical models,'' \emph{Latin American Journal of Central Banking}, vol.~4, no.~2, p. 100087, 2023. [Online]. Available: \url{https://www.sciencedirect.com/science/article/pii/S2666143823000042}
\BIBentrySTDinterwordspacing

\bibitem{GAN2020119928}
\BIBentryALTinterwordspacing
L.~Gan, H.~Wang, and Z.~Yang, ``Machine learning solutions to challenges in finance: An application to the pricing of financial products,'' \emph{Technological Forecasting and Social Change}, vol. 153, p. 119928, 2020. [Online]. Available: \url{https://www.sciencedirect.com/science/article/pii/S0040162519312399}
\BIBentrySTDinterwordspacing

\bibitem{ZENG2019376}
\BIBentryALTinterwordspacing
Y.~Zeng and D.~Klabjan, ``Online adaptive machine learning based algorithm for implied volatility surface modeling,'' \emph{Knowledge-Based Systems}, vol. 163, pp. 376--391, 2019. [Online]. Available: \url{https://www.sciencedirect.com/science/article/pii/S0950705118304350}
\BIBentrySTDinterwordspacing

\bibitem{9912385}
Y.~Xie, G.~Liu, C.~Yan, C.~Jiang, M.~Zhou, and M.~Li, ``Learning transactional behavioral representations for credit card fraud detection,'' \emph{IEEE Transactions on Neural Networks and Learning Systems}, vol.~35, no.~4, pp. 5735--5748, 2024.

\bibitem{9744717}
Y.~Xie, G.~Liu, C.~Yan, C.~Jiang, and M.~Zhou, ``Time-aware attention-based gated network for credit card fraud detection by extracting transactional behaviors,'' \emph{IEEE Transactions on Computational Social Systems}, vol.~10, no.~3, pp. 1004--1016, 2023.

\bibitem{10063201}
H.~Zhu, M.~Zhou, G.~Liu, Y.~Xie, S.~Liu, and C.~Guo, ``Nus: Noisy-sample-removed undersampling scheme for imbalanced classification and application to credit card fraud detection,'' \emph{IEEE Transactions on Computational Social Systems}, vol.~11, no.~2, pp. 1793--1804, 2024.

\bibitem{10319418}
Z.~Yi, X.~Cao, Z.~Chen, and S.~Li, ``Artificial intelligence in accounting and finance: Challenges and opportunities,'' \emph{IEEE Access}, vol.~11, pp. 129\,100--129\,123, 2023.

\bibitem{9738619}
Z.~D. Akşehir and E.~Kiliç, ``How to handle data imbalance and feature selection problems in cnn-based stock price forecasting,'' \emph{IEEE Access}, vol.~10, pp. 31\,297--31\,305, 2022.

\bibitem{10433778}
S.~Guha, F.~A. Khan, J.~Stoyanovich, and S.~Schelter, ``Automated data cleaning can hurt fairness in machine learning-based decision making,'' \emph{IEEE Transactions on Knowledge and Data Engineering}, vol.~36, no.~12, pp. 7368--7379, 2024.

\bibitem{THAKKAR2021114800}
\BIBentryALTinterwordspacing
A.~Thakkar and K.~Chaudhari, ``A comprehensive survey on deep neural networks for stock market: The need, challenges, and future directions,'' \emph{Expert Systems with Applications}, vol. 177, p. 114800, 2021. [Online]. Available: \url{https://www.sciencedirect.com/science/article/pii/S0957417421002414}
\BIBentrySTDinterwordspacing

\bibitem{ugwuishiwu2020overview}
\BIBentryALTinterwordspacing
C.~Ugwuishiwu, U.~Orji, C.~Ugwu, and C.~Asogwa, ``An overview of quantum cryptography and shor’s algorithm,'' \emph{Int. J. Adv. Trends Comput. Sci. Eng}, vol.~9, no.~5, 2020. [Online]. Available: \url{https://citeseerx.ist.psu.edu/document?repid=rep1&type=pdf&doi=c4c3ad4aef68f3970d187fec0f13755471579018}
\BIBentrySTDinterwordspacing

\bibitem{Nagata2017}
\BIBentryALTinterwordspacing
K.~Nagata, T.~Nakamura, and A.~Farouk, ``Quantum cryptography based on the deutsch-jozsa algorithm,'' \emph{International Journal of Theoretical Physics}, vol.~56, no.~9, pp. 2887--2897, Sep 2017. [Online]. Available: \url{https://doi.org/10.1007/s10773-017-3456-x}
\BIBentrySTDinterwordspacing

\bibitem{Du2022}
\BIBentryALTinterwordspacing
Y.~Du, T.~Huang, S.~You, M.-H. Hsieh, and D.~Tao, ``Quantum circuit architecture search for variational quantum algorithms,'' \emph{npj Quantum Information}, vol.~8, no.~1, p.~62, May 2022. [Online]. Available: \url{https://doi.org/10.1038/s41534-022-00570-y}
\BIBentrySTDinterwordspacing

\bibitem{Beer2020}
\BIBentryALTinterwordspacing
K.~Beer, D.~Bondarenko, T.~Farrelly, T.~J. Osborne, R.~Salzmann, D.~Scheiermann, and R.~Wolf, ``Training deep quantum neural networks,'' \emph{Nature Communications}, vol.~11, no.~1, p. 808, 2020. [Online]. Available: \url{https://doi.org/10.1038/s41467-020-14454-2}
\BIBentrySTDinterwordspacing

\bibitem{Orus2019}
\BIBentryALTinterwordspacing
R.~Orús, S.~Mugel, and E.~Lizaso, ``Quantum computing for finance: Overview and prospects,'' \emph{Reviews in Physics}, vol.~4, p. 100028, 2019. [Online]. Available: \url{https://www.sciencedirect.com/science/article/pii/S2405428318300571}
\BIBentrySTDinterwordspacing

\bibitem{NASTASIUK20151998}
\BIBentryALTinterwordspacing
V.~Nastasiuk, ``Fisher information and quantum potential well model for finance,'' \emph{Physics Letters A}, vol. 379, no.~36, pp. 1998--2000, 2015. [Online]. Available: \url{https://www.sciencedirect.com/science/article/pii/S037596011500571X}
\BIBentrySTDinterwordspacing

\bibitem{PAQUET2022116583}
\BIBentryALTinterwordspacing
E.~Paquet and F.~Soleymani, ``Quantumleap: Hybrid quantum neural network for financial predictions,'' \emph{Expert Systems with Applications}, vol. 195, p. 116583, 2022. [Online]. Available: \url{https://www.sciencedirect.com/science/article/pii/S0957417422000720}
\BIBentrySTDinterwordspacing

\bibitem{10545170}
H.~Mustafa and P.~Jain, ``Quantum graph neural networks for portfolio optimization in complex financial markets, a novel approach,'' in \emph{2024 International Conference on Trends in Quantum Computing and Emerging Business Technologies}, 2024, pp. 1--5.

\bibitem{PAPOUSKOVA201933}
\BIBentryALTinterwordspacing
M.~Papouskova and P.~Hajek, ``Two-stage consumer credit risk modelling using heterogeneous ensemble learning,'' \emph{Decision Support Systems}, vol. 118, pp. 33--45, 2019. [Online]. Available: \url{https://www.sciencedirect.com/science/article/pii/S0167923619300028}
\BIBentrySTDinterwordspacing

\bibitem{ai5040101}
\BIBentryALTinterwordspacing
E.~Mienye, N.~Jere, G.~Obaido, I.~D. Mienye, and K.~Aruleba, ``Deep learning in finance: A survey of applications and techniques,'' \emph{AI}, vol.~5, no.~4, pp. 2066--2091, 2024. [Online]. Available: \url{https://www.mdpi.com/2673-2688/5/4/101}
\BIBentrySTDinterwordspacing

\bibitem{McClean2018}
\BIBentryALTinterwordspacing
J.~R. McClean, S.~Boixo, V.~N. Smelyanskiy, R.~Babbush, and H.~Neven, ``Barren plateaus in quantum neural network training landscapes,'' \emph{Nature Communications}, vol.~9, no.~1, p. 4812, 2018. [Online]. Available: \url{https://doi.org/10.1038/s41467-018-07090-4}
\BIBentrySTDinterwordspacing

\bibitem{PhysRevLett.128.180505}
\BIBentryALTinterwordspacing
K.~Sharma, M.~Cerezo, L.~Cincio, and P.~J. Coles, ``Trainability of dissipative perceptron-based quantum neural networks,'' \emph{Phys. Rev. Lett.}, vol. 128, p. 180505, May 2022. [Online]. Available: \url{https://link.aps.org/doi/10.1103/PhysRevLett.128.180505}
\BIBentrySTDinterwordspacing

\bibitem{FraudDetection2022}
E.~Esenogho, I.~D. Mienye, T.~G. Swart, K.~Aruleba, and G.~Obaido, ``A neural network ensemble with feature engineering for improved credit card fraud detection,'' \emph{IEEE Access}, vol.~10, pp. 16\,400--16\,407, 2022.

\bibitem{CreditScoring2021}
S.~Ebiaredoh-Mienye, E.~Esenogho, and T.~Swart, ``Artificial neural network technique for improving prediction of credit card default: A stacked sparse autoencoder approach,'' \emph{International Journal of Electrical and Computer Engineering}, vol.~11, pp. 4392--4402, 10 2021.

\bibitem{ZHANG201865}
\BIBentryALTinterwordspacing
T.~ZHANG, W.~ZHANG, W.~XU, and H.~HAO, ``Multiple instance learning for credit risk assessment with transaction data,'' \emph{Knowledge-Based Systems}, vol. 161, pp. 65--77, 2018. [Online]. Available: \url{https://www.sciencedirect.com/science/article/pii/S0950705118303800}
\BIBentrySTDinterwordspacing

\bibitem{FeatureSelection2022}
\BIBentryALTinterwordspacing
S.~A. Ebiaredoh-Mienye, T.~G. Swart, E.~Esenogho, and I.~D. Mienye, ``A machine learning method with filter-based feature selection for improved prediction of chronic kidney disease,'' \emph{Bioengineering}, vol.~9, no.~8, 2022. [Online]. Available: \url{https://www.mdpi.com/2306-5354/9/8/350}
\BIBentrySTDinterwordspacing

\bibitem{DiseaseDetection2020}
\BIBentryALTinterwordspacing
S.~A. Ebiaredoh-Mienye, E.~Esenogho, and T.~G. Swart, ``Integrating enhanced sparse autoencoder-based artificial neural network technique and softmax regression for medical diagnosis,'' \emph{Electronics}, vol.~9, no.~11, 2020. [Online]. Available: \url{https://www.mdpi.com/2079-9292/9/11/1963}
\BIBentrySTDinterwordspacing

\bibitem{innan2024lepqnnloaneligibilityprediction}
\BIBentryALTinterwordspacing
N.~Innan, A.~Marchisio, M.~Bennai, and M.~Shafique, ``Lep-qnn: Loan eligibility prediction using quantum neural networks,'' 2024. [Online]. Available: \url{https://arxiv.org/abs/2412.03158}
\BIBentrySTDinterwordspacing

\bibitem{Innan_2024}
\BIBentryALTinterwordspacing
N.~Innan, A.~Sawaika, A.~Dhor, S.~Dutta, S.~Thota, H.~Gokal, N.~Patel, M.~A.-Z. Khan, I.~Theodonis, and M.~Bennai, ``Financial fraud detection using quantum graph neural networks,'' \emph{Quantum Machine Intelligence}, vol.~6, no.~1, Feb. 2024. [Online]. Available: \url{http://dx.doi.org/10.1007/s42484-024-00143-6}
\BIBentrySTDinterwordspacing

\bibitem{sakuma2022applicationdeepquantumneural}
\BIBentryALTinterwordspacing
T.~Sakuma, ``Application of deep quantum neural networks to finance,'' 2022. [Online]. Available: \url{https://arxiv.org/abs/2011.07319}
\BIBentrySTDinterwordspacing

\bibitem{math12091391}
\BIBentryALTinterwordspacing
N.~Schetakis, D.~Aghamalyan, M.~Boguslavsky, A.~Rees, M.~Rakotomalala, and P.~R. Griffin, ``Quantum machine learning for credit scoring,'' \emph{Mathematics}, vol.~12, no.~9, 2024. [Online]. Available: \url{https://www.mdpi.com/2227-7390/12/9/1391}
\BIBentrySTDinterwordspacing

\bibitem{beach2024creditcardfraud}
\BIBentryALTinterwordspacing
J.~Beach, ``Credit card fraud,'' 2024, accessed: 2024-10-25. [Online]. Available: \url{https://www.kaggle.com/datasets/joebeachcapital/credit-card-fraud}
\BIBentrySTDinterwordspacing

\bibitem{zohaib2024eligibilityprediction}
\BIBentryALTinterwordspacing
D.~Zohaib, ``Eligibility prediction for loan,'' 2024, accessed: 2024-10-25. [Online]. Available: \url{https://www.kaggle.com/datasets/devzohaib/eligibility-prediction-for-loan}
\BIBentrySTDinterwordspacing

\bibitem{RB202135}
\BIBentryALTinterwordspacing
A.~RB and S.~K. KR, ``Credit card fraud detection using artificial neural network,'' \emph{Global Transitions Proceedings}, vol.~2, no.~1, pp. 35--41, 2021, 1st International Conference on Advances in Information, Computing and Trends in Data Engineering (AICDE - 2020). [Online]. Available: \url{https://www.sciencedirect.com/science/article/pii/S2666285X21000066}
\BIBentrySTDinterwordspacing

\bibitem{haque2024bankloanpredictionusing}
\BIBentryALTinterwordspacing
F.~M.~A. Haque and M.~M. Hassan, ``Bank loan prediction using machine learning techniques,'' 2024. [Online]. Available: \url{https://arxiv.org/abs/2410.08886}
\BIBentrySTDinterwordspacing

\bibitem{9259940}
T.~Giurgica-Tiron, Y.~Hindy, R.~LaRose, A.~Mari, and W.~J. Zeng, ``Digital zero noise extrapolation for quantum error mitigation,'' in \emph{2020 IEEE International Conference on Quantum Computing and Engineering (QCE)}, 2020, pp. 306--316.

\end{thebibliography}

\vfill

\end{document}